\documentclass[aps,pre,twocolumn,showpacs,superscriptaddress,amsmath,amssymb,floatfix]{revtex4}
\usepackage{graphicx}
\usepackage{dcolumn}
\usepackage{bm}
\begin{document}

\title{The local structure factor near an interface; Beyond extended Capillary-Wave models}

\author{A.O.\ Parry}
\affiliation{Department of Mathematics, Imperial College London, London SW7 2BZ, United Kingdom}

\author{C.\ Rasc\'{o}n}
\affiliation{GISC, Departamento de Matem\'aticas, Universidad Carlos III de Madrid, 28911 Legan\'es, Madrid, Spain}

\author{R.\ Evans}
\affiliation{HH Wills Physics Laboratory, University of Bristol, Bristol BS8 1TL, United Kingdom}

\begin{abstract}
We investigate the local structure factor $S(z;q)$ at a free liquid-gas interface in systems with short-ranged intermolecular forces and determine the corrections to the leading-order, capillary-wave-like, Goldstone mode divergence of $S(z;q)$ known to occur for parallel wavevectors $q\to 0$. We show from explicit solution of the inhomogeneous Ornstein-Zernike equation that for distances $z$ far from the interface, where the profile decays exponentially, $S(z;q)$ splits unambiguously into bulk and interfacial contributions. On each side of the interface, the interfacial contributions can be characterised by distinct liquid and gas wavevector dependent surface tensions, $\sigma_l(q)$ and $\sigma_g(q)$, which are determined solely by the {\it{bulk}} two-body and three-body direct correlation functions. At high temperatures, the wavevector dependence simplifies and is determined almost entirely by the appropriate bulk structure factor, leading to positive rigidity coefficients. Our predictions are confirmed by explicit calculation of $S(z;q)$ within square-gradient theory and the Sullivan model. The results for the latter predict a striking temperature dependence for $\sigma_l(q)$ and $\sigma_g(q)$, and have implications for fluctuation effects. Our results account quantitatively for the findings of a recent very extensive simulation study by H\"ofling and Dietrich of the total structure factor in the interfacial region, in a system with a cut-off Lennard-Jones potential, in sharp contrast to extended Capillary-Wave models which failed completely to describe the simulation results.
\end{abstract}


\maketitle

\section{Introduction}

Is there such a thing as a wavevector dependent surface tension $\sigma(q)$ which controls the (free) energy cost of the fluctuations of a liquid-gas interface? If so, how does one define, predict, and measure $\sigma(q)$? Does it increase or decrease with parallel wavevector $q$? These questions have been raised by a number of authors over the last few decades, using different approaches, with conflicting conclusions drawn. In the present paper, we show that the concept of "a" wavevector dependent surface tension, suitably generalized, can be placed firmly within the framework of the modern microscopic theory of inhomogeneous fluids, including density functional theory (DFT). This can be done by focusing on the properties of the local structure factor $S(z;q)$, where $z$ is the distance normal to the interface. However, in doing this, we are led to a number of somewhat surprising conclusions including that there must be not one but two wavevector dependent surface tensions $\sigma_l(q)$ and $\sigma_g(q)$, which characterise the properties of $S(z;q)$ on the liquid and gas sides of the interface, respectively. We show that these separate tensions $\sigma_l(q)$ and $\sigma_g(q)$ reconcile fully theory with the results of recent extensive simulation studies, which were not consistent with the predictions of extended capillary-wave models. Indeed, our present microscopic theory highlights a number of inadequacies of extended capillary-wave models, particularly when one attempts to derive these from DFT.\\ 

The role of the surface tension in determining fluctuations at large wavelengths, $q\to 0$, is well understood. At lengthscales much larger than the bulk correlation length, equilibrium density fluctuations of a near planar liquid-gas interface, pinned by a weak gravitational field, are dominated by undulations in the local interface "height" $\ell({\bf{x}})$, with ${\bf x}=(x,y)$. At these scales, the energy cost of small amplitude height fluctuations is described by the classical Capillary-Wave Hamiltonian \cite{BLS1965,Weeks1977,Bedeaux1985}
\begin{equation}
H[\ell]\,=\int\!\!d{\bf{x}}\;\left\{\;\frac{\sigma}{2}(\nabla \ell)^2+\frac{mg\Delta\rho}{2}\,\ell^2\,\right\}
\label{CWreal}
\end{equation}
where $\sigma$ is the planar surface tension, $m$ is the molecular mass, $g$ is the gravitational acceleration and $\Delta\rho=\rho_l\!-\!\rho_g$, is the difference in bulk number densities of coexisting liquid and gas. In terms of the Fourier amplitudes of $\ell({\bf{x}})$, this reads
\begin{equation}
H[\ell]\;=\;\frac{1}{2A}\,\sum_q\;\left(mg\Delta\rho+\sigma q^2\right)\,
|\tilde\ell(q)|^2
\label{CWFourier}
\end{equation}
where the sum runs from $q_{min}\approx 1/L_\parallel$, set by the interfacial area $A=L_\parallel^2$, to a high-momentum cut-off $q_{max}$, of order the inverse bulk correlation length, beyond which the classical Capillary-Wave description breaks down. Thus, equipartition implies that
\begin{equation}
 \beta\,\left\langle|\tilde\ell(q)|^2\right\rangle\;=\;\frac{A}{\;mg\Delta\rho+\sigma q^2\,}
\label{CW}
\end{equation}
where the LHS corresponds to the Fourier transform of the height-height correlation function $\langle \ell({\bf{x}})\ell(0)\rangle$. Here, $\beta=1/k_B T$. Note that the height-height correlations exhibit a Goldstone mode when $g=0^+$, due to the fact that translations in the interfacial height then cost zero energy.  In three dimensions, this means that the r.m.s.~interfacial width $\xi_\perp=\sqrt{\langle \ell^2\rangle}$, is broadened weakly by fluctuations: $\xi_\perp^2=(2\pi\beta\sigma)^{-1}\ln  (\xi_\parallel q_{max})$, where $\xi_\parallel=\sqrt{\sigma/mg\Delta\rho}$ is the parallel correlation or capillary length. A similar result applies in zero field, in which case the role of $\xi_\parallel$ is replaced by $L_\parallel$. This simple interpretation of interfacial fluctuations, akin to the thermal excitations of a drumskin under tension, is in keeping with exact sum-rules requirements for the density-density correlation function $G({\bf{r}},{\bf{r}}')=\langle\hat \rho({\bf{r}})\hat\rho({\bf{r}}')\rangle\!-\!\rho(z)\rho(z')$, where $\hat\rho({\bf{r}})=\sum_i\delta({\bf{r}}\!-\!{\bf{r}}_i)$ is the usual density operator, and $\rho(z)=\langle \hat \rho({\bf{r}})\rangle$ is the equilibrium density profile. In a pioneering study \cite{Wertheim1976}, Wertheim showed using expressions such as the exact Triezenberg-Zwanzig result \cite{Triezenberg1972} for the surface tension, that, at small $q$,
\begin{equation}
\beta\, G(z,z';q)\;\approx\; \frac{\rho'(z)\rho'(z')}{\;mg\Delta\rho+\sigma q^2\,}
\label{GGM}
\end{equation}
where $\rho'(z)=d\rho(z)/dz$ and $G(z,z';q)$ is the parallel Fourier transform of $G({\bf{r}},{\bf{r}}')$. The dependence on $\rho'(z)$ in this expression is completely consistent with the Capillary-Wave theory expectation that long wavelength interfacial fluctuations merely translate the density profile. Integration implies that the local structure factor $S(z;q)\!=\!\!\int\! dz'\,G(z,z';q)$ has the form
\begin{equation}
\beta\, S(z;q)\;\approx\; \frac{\rho'(z)\,\Delta\rho}{\;mg\Delta\rho+\sigma q^2\,}
\label{SGM}
\end{equation}
where the $z$ axis is chosen so that $\rho'(z)>0$. Indeed, the RHS of (\ref{SGM}) is the \textit{exact} expression within Capillary-Wave theory when one adopts a step function for the density operator or intrinsic profile (see Appendix). This observation suggests that it may be through the local structure factor, rather than the more complicated two point function $G(z,z';q)$, that  position dependent density response functions display the simplest dependence on $q$. Obviously, the expression (\ref{SGM}) does not describe the full behaviour in a real fluid since, far from the interfacial region, where $\rho'(z)\to 0$, we must have that $S(z;q)$ approaches the value of the bulk structure factor in the appropriate liquid or gas phase. Nevertheless, in the interfacial region and at least for long-wavelengths $q<q_{max}$, there is pleasing consistency between the microscopic sum rules and mesoscopic Capillary-Wave theory. We mention also that prediction (\ref{CW}) for the height-height correlation function has been confirmed conclusively in experimental studies of colloid-polymer interfaces, where the interfacial fluctuations can be imaged directly \cite{Aarts2004}.\\

\begin{figure}[b]
\includegraphics[width=\columnwidth]{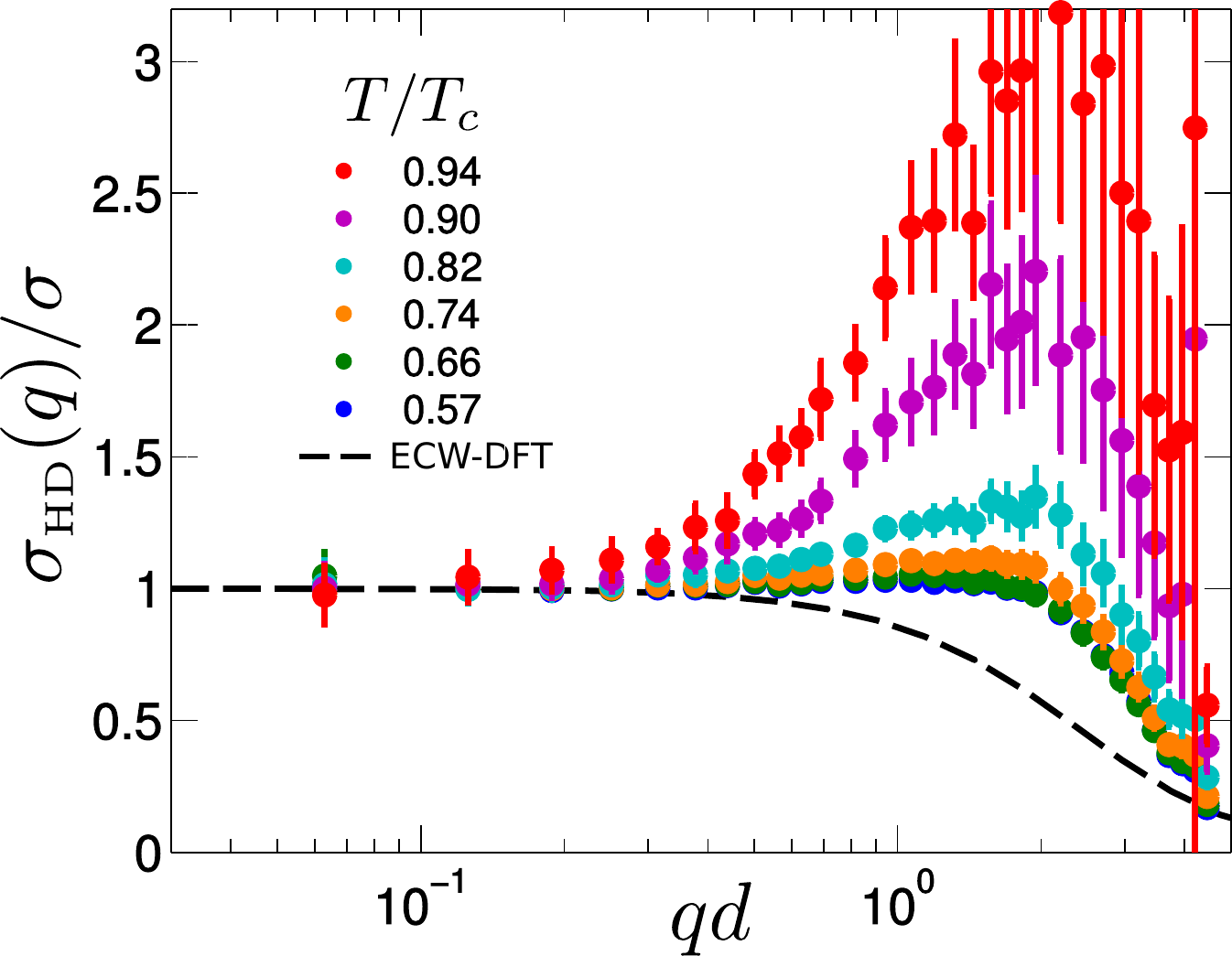}
\caption{\label{Fig1} H\"ofling-Dietrich (HD) simulation results for the wavevector dependent tension for a cut-off Lennard-Jones potential at different reduced temperatures. The tension $\sigma_\textup{\tiny HD}(q)$ is defined from the total structure factor $S(q)=S^{ex}(q)+S^{bg}(q)$, where $S^{ex}(q)=(\Delta\rho)^2/\beta\sigma_\textup{\tiny HD}(q)q^2$ is the excess contribution and $S^{bg}(q)$ the background signal which is a weighted combination of the bulk liquid and gas structure factors (see later). The dashed line is the wavevector dependent tension $\sigma_\textup{\tiny ECW}^\textup{\tiny DFT}(q)$ for the extended Capillary-Wave model as derived using DFT. Here, $d\equiv\sigma_\textup{LJ}$ is the usual Lennard-Jones  diameter and $T_c$ is the bulk critical temperature. Adapted from Fig. 3(b) of Ref. \cite{Hofling2015}.}
\end{figure}

Over the last 20 years, concerted efforts have been made to understand the nature of fluctuations and the structure of $G(z,z';q)$ and $S(z;q)$ occurring for larger wavevectors \cite{Hofling2015,Zia1985,Rochin1991,Blokhuis1993,Napiorkowski1993,Parry1994,Mecke1999,Blokhuis1999,Fradin2000,Mora2003,Vink2005,Blokhuis2008,Blokhuis2009,Chacon2014}. By far the most common approach to this problem has been to assume that one can extend the Capillary-Wave Hamiltonian (\ref{CWFourier}) by relinquishing the upper momentum cut-off $q_{max}$ and, instead, introduce a wavevector dependent surface tension, so that 
\begin{equation}
H[\ell]\;=\;\frac{1}{2A}\sum_q \sigma_\textsc{\tiny ECW}(q)\, q^2 |\tilde\ell(q)|^2
\end{equation}
where, for convenience, we have now set $g=0^+$, but kept the area $A=L_\parallel^2$ finite.  For systems with short-ranged forces, one may expand the assumed wavevector dependent tension $\sigma_\textsc{\tiny ECW}(q)=\sigma+\mathcal{K}_\textsc{\tiny ECW} q^2+\dots$ and thus define a rigidity $\mathcal{K}_\textsc{\tiny ECW}$, which is commonly regarded as arising from curvature contributions to the extended Capillary-Wave Hamiltonian $H[\ell]$, analogous to those occurring in the mesoscopic theory of membranes \cite{Zia1985}. There are no significant consequences for fluctuation effects associated with a rigidity; for example, it does not alter the basic dependence of $\xi_\perp$ on $\xi_\parallel$. Nevertheless, it has been hoped that one might use the concept of a wavevector dependent tension to explain and predict structural properties in the interfacial region at higher wavevectors $q$, beyond the limits of traditional Capillary-Wave theory, perhaps leading to the enhancement or suppression of fluctuation effects. In particular, from the modified expression for the height-height correlation function
\begin{equation}
\beta\,\left\langle|\tilde\ell(q)|^2\right\rangle\;=\;\frac{A}{\;\sigma_\textsc{\tiny ECW}(q)\, q^2\,}
\label{ECW1}
\end{equation}
one might surmise that one could predict the properties of $G$ and $S$ at higher wavevectors. We mention that, for systems with dispersion forces, there are necessarily logarithmic corrections to the rigidity arising from the 2D Fourier transform of the $r^{-6}$ tail of the intermolecular potential \cite{Napiorkowski1993}. This is something we will return to at the end of our paper. For now, we concentrate on systems with short-ranged forces, such as cut-off Lennard-Jones or square-well potentials, commonly used in simulation studies. Such systems exhibit the conceptual difficulties inherent in this subject.\\

Despite its plausibility, this simple extension of Capillary-Wave theory is stricken with problems. Attempts to derive $\mathcal{K}_\textsc{\tiny ECW}$ using the modern microscopic theory of inhomogeneous fluids show that its value, unlike that of the surface tension $\sigma$, is dependent on the definition of the interface position $\ell({\bf{x}})$. Indeed even the {\it{sign}} of the rigidity has been contentious. We have discussed these problems in detail in two recent papers \cite{Parry2014,Parry2015}. For example, beyond the long wavelength limit, interfacial and density fluctuations are no longer related by a simple translation of the density profile, so that even if one could derive (\ref{ECW1}) one could not infer the structure of $G(z,z';q)$ and $S(z;q)$ without very detailed additional information about density fluctuations \cite{Parry2014}. This makes linking with experiments near impossible. Relating $\langle |\tilde\ell(q)|^2\rangle$, $G(z,z';q)$ and $S(z;q)$ is further complicated when one allows for a realistic asymmetry between the liquid and gas phases, i.e. different bulk correlation lengths, which introduces the problem of splitting these three functions consistently into background and interfacial contributions \cite{Parry2015}.\\

These issues have recently come to a head in simulation studies of the interfacial region in a system with cut-off Lennard-Jones forces. In an extensive molecular dynamics study, involving up to $5\times 10^5$ particles, H\"ofling and Dietrich \cite{Hofling2015} extracted an effective wavevector dependent surface tension $\sigma_\textsc{\tiny HD}(q)$ from measurement of the total structure factor $S(q)\!=\!\int\!dz\,S(z;q)$. They find that $\sigma_\textsc{\tiny HD}(q)$ increases with $q$ before exhibiting a maximum and then a rapid decrease (see Fig.~\ref{Fig1}). Moreover $\sigma_\textsc{\tiny HD}(q)/\sigma$ depends strongly on temperature, the maximum being most pronounced at large values of $T/T_c$, where $T_c$ is the bulk critical temperature. As HD point out, this behaviour is at odds with extended capillary-wave theory, since the DFT prediction for $\mathcal{K}_\textsc{\tiny ECW}^\textsc{\tiny DFT}$ is {\it{negative}} so that $\sigma_\textsc{\tiny ECW}(q)$ simply decreases with increasing $q$ and is essential $T$ independent (See Fig.~\ref{Fig1}). \\

In this paper, we show that these conceptual issues may be overcome or avoided by determining directly $S(z;q)$ using microscopic theory. It turns out that the problems of explaining the simulation data for $S(q)$, and of defining a rigidity and determining its sign, do not lie with DFT but rather in trying to marry DFT with the extended Capillary-Wave expression (\ref{ECW1}). For wavevectors larger than the inverse bulk correlation lengths, which must be accessed in order to expose the corrections to the Goldstone mode term in $S(z;q)$, one also encounters the microscopic structure of the interfacial region and thus the different bulk properties of the liquid and gas phases. We shall see that this cannot be incorporated using the hypothesis (\ref{ECW1}), meaning it is not appropriate to model the interface as a fluctuating drum-skin at these larger wavevectors - at least not as a tool for understanding $S(z;q)$. However, the concept of a wavevector dependent tension {\it{does}} remain useful provided one defines it through $S(z;q)$ and recognises that there are {\it separate} tensions $\sigma_l(q)$ and $\sigma_g(q)$, on the liquid and gas sides of the interface. In Sec.~II we show that, for systems with short-ranged forces, these quantities can be identified unambiguously from the asymptotic decay, $|z|\to\infty$, of $S(z;q)$ , and are determined by the two-body and, to a lesser extent, three-body direct correlation functions of the appropriate {\it{bulk}} liquid or gas phases. In Sec.~III we determine $\sigma_l(q)$ and $\sigma_g(q)$ explicitly within square-gradient and the Sullivan model DFTs. These reveal a sensitive dependence on temperature, very different from the expectations of extended Capillary-Wave models. Sec.~IV shows how using our approach one can explain quantitatively the H\"ofling-Dietrich simulation results over the full range of temperatures. We draw conclusions in Sec. V.\\

\section{The "far" behaviour of $S(z;q)$.} 

\subsection{Exact Sum-rules and DFT}

Within DFT, the equilibrium density profile of an inhomogeneous fluid in an external field $V({\bf{r}})$ is found from minimization of the Grand potential functional \cite{Evans1979,Rowlinson1982,Evans1990, Hansen2006}
\begin{equation}
\Omega_V[\rho]\;=\;\mathcal{F}[\rho]-\int\!\!d{\bf{r}}\;u(\bf{r})\rho({\bf{r}})
\label{Omegagen}
\end{equation}
where $\mathcal{F}[\rho]$ is the intrinsic Helmholtz free-energy functional and $u({\bf r})=\mu\!-\!V(\bf{r})$, with $\mu$ the chemical potential. Thus, for a potential $V(z)$, the equilibrium profile $\rho({\bf{r}})=\rho(z)$ satisfies
\begin{equation}
\frac{\delta\Omega_V[\rho]}{\delta \rho({\bf{r}})}\,=\,0
\label{ELgen}
\end{equation}
subject to suitable boundary conditions, which impose a liquid density as $z\to\infty$ and gas as $z\to-\infty$. Complementing this, there is a hierarchy of distribution functions which are generated by successive functional differentiation of the equilibrium grand potential $\Omega=\Omega_V[\rho({\bf{r}})]$ with respect to $u(\bf{r})$ at fixed temperature $T$ . Thus, quite generally
\begin{equation}
\frac{\delta\Omega}{\delta u({\bf{r}})}\,=\,-\rho({\bf{r}})
\end{equation}
and
\begin{equation}
\beta^{-1}\,\frac{\delta\rho({\bf{r}})}{\delta u({\bf{r}}')}\,=\,G({\bf{r}},{\bf{r}}')
\end{equation}
These, in turn, lead to a number of exact sum-rules which involve the moments  $G_{2n}(z,z')$ which are defined from the parallel Fourier transform of the density-density correlation function
\begin{equation}
G(z,z';q)\;=\int\!\! d{\bf x}\;G({\bf r},{\bf r}')\,e^{i{\bf{q}}\cdot{\bf x}}
\label{FourierG}
\end{equation}
via the expansion
\begin{equation}
G(z,z';q)\,=\,G_0(z,z')\,+\,G_2(z,z')\,q^2\,+\,\cdots
\end{equation}
In (\ref{FourierG}) ${\bf x}$ is the parallel separation between the particles. For example, the zeroth moment satisfies \cite{Evans1979,Rowlinson1982,Henderson1992}
\begin{equation}
\rho'(z)\,=\,-\beta \int\!\! dz'\;V ' (z')\,G_0(z,z')
\label{Y1}
\end{equation}
with $V'(z)=d V(z)/d z$, equivalent to the Yvon equation \cite{Wertheim1976,Evans1979}, while the second-moment satisfies
\begin{equation}
\sigma\;=\,-\beta\int\!\!\int \!dz\,dz'\;\,V '(z')\,V'(z)\,G_2(z,z')
\label{sum2}
\end{equation}
which is equivalent \cite{Wertheim1976} to the Triezenberg-Zwanzig equation for the tension $\sigma$ \cite{Triezenberg1972}. As shown by Wertheim \cite{Wertheim1976}, for an interface in a gravitational field  $V(z)\!=\!-mgz$, a spectral analysis of the sum-rules leads to the expression (\ref{GGM}) for the correlation function, and hence to (\ref{SGM}) for the local structure factor.  We anticipate that (\ref{GGM}) and (\ref{SGM}) are valid at long wavelengths and $z$, $z'$ sufficiently close to the interface. There are also sum-rules associated with the direct correlation function. Within DFT, this is identified as
\begin{equation}
C({\bf{r}},{\bf{r}}')\;=\;\beta\,\frac{\delta^2\mathcal{F}[\rho]}{\,\delta \rho({\bf{r}}) \delta \rho({\bf{r}}')\,}
\label{Cdef}
\end{equation}
where the functional derivative is evaluated at the equilibrium  profile $\rho(z)$. Note that 
\begin{equation}
C({\bf{r}},{\bf{r}}')\;=\;\frac{\delta({\bf{r}}-{\bf{r}'})}{\rho({\bf{r}})}-c^{(2)}({\bf{r}},{\bf{r}}')
\end{equation}
where $c^{(2)}({\bf{r}},{\bf{r}}')$ is the usual Ornstein-Zernike pair direct correlation function of the inhomogeneous fluid \cite{Evans1979,Rowlinson1982,Hansen2006}.
For a planar interface it is again convenient to take the parallel Fourier transform
\begin{equation}
C(z,z';q)\;=\int\!d{\bf{x}}\;\,C({\bf{r}},{\bf{r}}')\,e^{i{\bf{q}}\cdot\bf{x}}
\label{CFourierdef}
\end{equation}
which, for systems with short-ranged forces, may also be expanded as
\begin{equation}
C(z,z';q)\;=\;C_0(z,z')\,+\,C_2(z,z')\,q^2\,+\,\cdots
\end{equation}
The direct-correlation function $C$ and pair correlation function $G$ are functional inverses which are related by the inhomogeneous Ornstein-Zernike equation
\begin{equation}
\int\! dz'' \;C(z,z'';q)\;G(z'',z';q)\;=\;\delta(z-z')
\label{OZeqn}
\end{equation}
allowing the sum-rules (\ref{Y1}) and (\ref{sum2}) to be re-expressed as
\begin{equation}
\beta\, V'(z)\;=\;-\int\! dz'\;\rho'(z')\,C_0(z,z')
\label{YvonC}
\end{equation}
and
\begin{equation}
\beta\,\sigma\;=\;\int\!\!\int\! dz\,dz'\;\,\rho'(z')\,\rho'(z)\, C_2(z,z')
\label{TZ}
\end{equation}
which {\it{is}} the Triezenberg-Zwanzig formula \cite{Triezenberg1972}. Thus, DFT provides a thermodynamically consistent route for determining the local structure factor,
\begin{equation}
S(z;q)\;=\,\int\! dz'\;G(z,z';q)
\end{equation}
from the solution of 
\begin{equation}
\int\! dz'\; C(z,z';q)\,S(z';q)\;=\;1
\label{OZS}
\end{equation}
which is simply the {\it{integrated}} inhomogeneous Ornstein-Zernike equation following from (\ref{OZeqn}). Many other exact sum-rules can be derived in this way and these have proved extremely useful in the study of fluid adsorption at planar walls, in capillary slits, and in other geometries \cite{Henderson1992}. We mention also that one may easily extend the hierarchy of distribution functions to include many-body direct and density correlation functions, which are related via generalizations of the Ornstein-Zernike equation \cite{Henderson1987}. \\

\subsection{Local closure of the Ornstein-Zernike equation from linear response}

Infinitely far from the interface, the pair correlation functions on the liquid and gas sides must take their bulk forms. In 3D these decay as
\begin{equation}
G_l(r)\,\propto\, \frac{\;e^{-\kappa_l r}\,}{r}
\hspace{.5cm}\textup{and}\hspace{.5cm}
G_g(r)\,\propto\, \frac{\;e^{-\kappa_g r}\,}{r}
\label{Gbulk}
\end{equation}
as $r\to\infty, $where we have now specialised explicitly to systems with short-ranged intermolecular forces. The inverse lengthscales determining the decay are $\kappa_l=1/\xi_l^T$ and $\kappa_g=1/\xi_g^T$, where $\xi_l^T$ and $\xi_g^T$ are the {\it{true}} correlation lengths of the bulk liquid and gas phases. In writing these expressions, we have assumed that the asymptotic decay of the correlation function is monotonic. This is always the case in the gas phase but, 
for the liquid, applies only for $T>T_\textsc{\tiny FW}$, where $T_\textsc{\tiny FW}$ is the temperature at which the Fisher-Widom line intersects the bulk coexistence curve \cite{Evans1993}. For $T<T_\textsc{\tiny FW}$, $G_l(r)$ decays as $e^{-\kappa_l r}\cos(\lambda r)/r $ The 3D Fourier transform of $G_b(r)$ defines the bulk structure factors
\begin{equation}
S_b(k)\equiv
\int\! d{\bf r}\;e^{ik\cdot {\bf{r}}}\;G_b(r) \,=\,\frac{S_b(0)}{\;1+\xi_b^2k^2+\cdots}
\end{equation}
where $S_b(0)=k_B T(\partial \rho_b/\partial \mu)_T$ is proportional to the bulk compressibility, and $\xi_b$ is the Ornstein-Zernike or second-moment correlation length of the bulk liquid ($b=l$) or gas ($b=g$). The inverse of the bulk structure factors identifies the Fourier transform of the bulk direct correlation functions
\begin{equation}
C_b(k)\;=\;\frac{1}{S_b(k)}
\label{CS}
\end{equation}
whose imaginary roots lying closest to the real axis determine the true bulk correlation lengths \cite{Evans1993}:
\begin{equation}
C_b(i\kappa_b)=0
\label{pole}
\end{equation}
The values of the second-moment correlation length $\xi_b$, and true correlation length $\xi_b^T$ are similar at high temperatures, and are, of course, identical in square gradient DFT. We note that the structure factor introduced here has dimensions $1/$volume, i.e. in bulk $S_b(k)\equiv G_b(k)$, the Fourier transform of the bulk density-density correlation function. The (dimensionless) static structure factor usually employed in liquid state theory \cite{Hansen2006} is $\rho_b^{-1}S_b(k)$, where $\rho_b$ is the number density. Equation (\ref{CS}) is equivalent to the standard bulk Ornstein-Zernike equation in Fourier space.\\

Next, let us consider the liquid-gas interface in such a system. At mean-field level or in dimensions $d>3$, $\rho'(z)\neq 0$ in zero external field and the profile will decay exponentially towards the bulk densities. Thus, deep in the liquid phase ($z\to\infty$), we can write
\begin{equation}
\rho(z)\;=\;\rho_l\;-\;\alpha_l\, e^{-\kappa_l z}\,+\,\mathcal{O}(e^{-2\kappa_l z})
\label{rhodecayl}
\end{equation}
while, for the gas ($z\to -\infty$),
\begin{equation}
\rho(z)\;=\;\rho_g\;+\;\alpha_g\,e^{-\kappa_g |z|}\,+\,\mathcal{O}(e^{-2\kappa_g |z|})
\label{rhodecayg}
\end{equation}
Note that the inverse lengthscales are the true correlation lengths of the bulk liquid and gas phases and, once again, we have assumed that $T>T_\textsc{\tiny FW}$, so that the decay of $\rho(z)$ is monotonic on both sides of the interface \cite{Evans1993}. These forms for the decay of $\rho(z)$ remain valid in $d=3$, beyond mean-field, although due to the influence of the capillary-wave broadening of the interface, one must impose that the interfacial area $L_\parallel^2$ remains finite. In this case, the exponential decays are appropriate for distances $|z|\gg\xi_\perp$ and one must allow for the amplitudes to be renormalized from their mean-field values. Capillary-Wave theory predicts that this renormalization induces a finite-size scaling dependence described by
\begin{equation}
\alpha_l(L_\parallel)\;\approx\; L_\parallel ^{\omega_l}
\hspace{.5cm}\textup{and}\hspace{.5cm}
\alpha_g(L_\parallel)\;\approx\; L_\parallel^{\omega_g}
\label{Scaling}
\end{equation} 
where $\omega_b\!=\!k_B T\kappa_b^2/4\pi\sigma$ is the dimensionless 'wetting' parameter that describes the strength of the fluctuations on the liquid ($b=l$) and the gas ($b=g$) side \cite{Henderson1992,Evans1989,Evans1992,Fernandez2013}.\\

We wish to determine how $S(z;q)$ decays towards its bulk liquid or gas limit as $|z|\to \infty$ at {\it{fixed}} $q>0$. On  the liquid side ($z\to \infty$), this decay must have the form
\begin{equation}
S(z;q)\;=\; S_l(q)\,+\,\frac{\rho'(z)\,\Delta \rho}{\;\beta\,\sigma_l(q)\, q^2\,}\,+\,\cdots
\label{Slargel}
\end{equation}
where $\sigma_l(q)$ is to be determined, and the higher-order terms decay faster than $e^{-\kappa_l z}$, which is the leading order decay of $\rho(z)$.
The $z$ dependence and the coefficient $\sigma_l(q)$ follow from the solution of the Ornstein-Zernike equation. To see this, we substitute (\ref{Slargel}) into (\ref{OZS}) to obtain
\begin{equation}
\int\!dz'\,C(z,z';q)\,S_l(q)\,+\,\frac{\Delta \rho}{\beta\sigma_l(q) q^2}\int\! dz'\,C(z,z';q)\,\rho'(z')=1
\end{equation}
 and then expand the direct correlation function about the bulk liquid density
\begin{equation}
\begin{split}
C(z,z';q)\; & = \; C_l(|z-z'|;q)\;+\\[.2cm]
& +\int\! dz''\;\, \frac{\,\delta C(z,z';q)\,}{\delta\rho(z'')}\,\delta\rho(z'')\,+\,\cdots
\end{split}
\label{Cexpansion}
\end{equation}
where the functional derivative is evaluated at $\rho(z)=\rho_l$, and we have defined $\delta \rho=\rho(z)\!-\!\rho_l$. The first term on the RHS of (\ref{Cexpansion}) is the parallel Fourier transform of the bulk liquid direct correlation function, and isotropy of the homogeneous fluid implies
\begin{equation}
\int\!dz'\;C_l(|z-z'|;q)\;=\;C_l(q)
\end{equation}
For large $z$, we therefore require that
\begin{equation}
\begin{split}
\int\!\!\!\int &dz'dz''\;\,\frac{\,\delta C(z,z';q)\,}{\delta\rho(z'')}\; \delta\rho(z'')\;=\\[.2cm]
&-\frac{\Delta\rho\, C_l(q)}{\;\beta\sigma_l(q) q^2\,}\int\!\! dz'\,C_l(|z-z'|;q)\,\rho'(z')
\end{split}
\end{equation}
where we have used (\ref{CS}). If we substitute $\rho'(z)\approx-\kappa_l\,\delta\rho(z)\propto e^{-\kappa_l z}$, we can see that both integrals are of order $e^{-\kappa_l z}$ and equating them determines 
\begin{equation}
q^2\beta\sigma_l(q)\;=\;\kappa_l\,\Delta\rho\; C_l(q)\;\frac{\displaystyle\int\! dz'\;\, C_l(|z-z'|;q) \,e^{-\kappa_l z'}}
{\displaystyle\int\!\!\!\int\! dz'\,dz''\;\frac{\delta C(z,z';q)}{\delta\rho(z'')} \,e^{-\kappa_l z''}}
\label{Gen0}
\end{equation}
 Note that the $z$ dependence in the numerator and denominator cancels since, in the bulk, $\frac{\delta C(z,z';q)}{\delta\rho(z'')}$ depends only on $|z\!-\!z'|$ and $|z\!-\!z''|$. The integrals exist since, away from the critical point, the bulk direct correlation function and its functional derivative decay over a short finite range, set by the finite range of the intermolecular potential \cite{Hansen2006}. Next, we note that the functional derivative in the denominator is related to the three-body direct correlation function defined as
\begin{equation}
C^{(3)}({\bf{r}}_1,{\bf{r}}_2,{\bf{r}}_3)\;=\;\frac{\,\delta C({\bf{r}}_1,{\bf{r}}_2)\,}{\delta \rho({\bf{r}}_3)}
\end{equation}
which, in turn, is related to the density-density-density correlation function by a many-body generalisation of the Ornstein-Zernike equation - see, for example, \cite{Henderson1987}. For our present purposes, we need only note that in the bulk liquid $C^{(3)}_l({\bf{r}}_1,{\bf{r}}_2,{\bf{r}}_3)$ depends only on three variables - the scalar distances $r_{12}$ and $r_{13}$, and also the angle between the vectors ${\bf{r}}_{12}$ and $\bf{r}_{13}$. This means that the double Fourier transform
\begin{equation}
\begin{split}
&C^{(3)}_l(q_{12};q_{13}^{z},q_{13}^{\parallel})=\\[.2cm]
&\;\;\int\!\!\!\int\! d{\bf{r}_{12}}\, d{\bf{r}_{13}}\;\,
e^{i({\bf{q}_{12}}\cdot{\bf{r}}_{12}+ {\bf{q}_{13}}\cdot{\bf{r}}_{13})}\;
C^{(3)}_l({\bf{r}}_{12},{\bf{r}}_{13})
\end{split}
\end{equation}
also depends on three variables. These can be taken to be the wavenumber $q_{12}$ of one Fourier transform and the wavevector $(q_{13}^{z},q_{13}^{\parallel})$ of the second as measured in convenient directions ($z$ and the parallel $XY$ plane, say). One can see that the functional derivative in the denominator of (\ref{Gen0}) is just the parallel Fourier transform of $C^{(3)}_l({\bf{r}}_1,{\bf{r}}_2,{\bf{r}}_3)$ at fixed $z_1$, $z_2$ and $z_3$, evaluated at $q_{12}=q$ and $q_{13}^\parallel=0$. Moreover, the weighted integrals over $e^{-\kappa_l z}$ are Bromwich integrals for the inverse Laplace transform obtained by analytic continuation into the complex plane. Making use of the isolated pole (\ref{pole}) defining the true correlation length, it follows that this cumbersome expression simplifies to
\begin{equation}
q^2\beta\sigma_l(q)\;=\,\kappa_l\,\Delta\rho\; \frac{C_l(q)\,C_l(q_*)}{\;C_l^{(3)}(q;i\kappa_l,0)\,}
\label{Gen}
\end{equation}
where
\begin{equation}
q_*\,=\;i\,\sqrt{\kappa_l^2-q^2}
\label{qstar}
\end{equation}
The imaginary value of $q_*$ at small $q$ is not a problem since $C_l(q)$ is an even function of $q$. Note that  the coefficient $\sigma_l(q)$ characterises the {\it{full}} wavevector dependence of the leading-order exponential decay of $S(z;q)$, and is not an expansion in $q$.\\
Equivalent results apply on the gas side where for {\it{fixed}} $q>0$ and $z\to -\infty$, the local structure factor decays as
\begin{equation}
S(z;q)\;=\; S_g(q)\;+\;\frac{\rho'(z)\,\Delta \rho}{\;\beta\,\sigma_g(q)\, q^2\,}\;+\;\cdots
\label{Slargeg}
\end{equation}
where, now, the higher-order terms decay faster than $e^{-\kappa_g|z|}$, and one determines that
\begin{equation}
q^2\,\beta\sigma_g(q)\;= -\;\kappa_g\,\Delta\rho\; \frac{C_g(q)\, C_g(q_*)}{\;C_g^{(3)}(q;i\kappa_g,0)\,}
\label{Geng}
\end{equation}
where now $q_*= i \sqrt{\kappa_g^2-q^2}$ and the minus sign arises directly from the decay of the density profile on the gas side. We emphasise that in (\ref{Gen}) and (\ref{Geng}) the direct correlation functions refer to the appropriate {\it{bulk}}. These equations are formally exact expressions for the two wavevector dependent tensions $\sigma_l(q)$ and $\sigma_g(q)$ describing the "far" behaviour of $S(z;q)$, that is the limit $|z|\to\infty$ at {\it{fixed}} $q$, in systems with short-ranged forces, provided that at large distances $\rho(z)$ decays exponentially. We also deliberately avoid the very close proximity of $T_c$. \\

Note that, in the limit $q\to 0$, the coefficients $\sigma_b(q)$ tend to a non-zero limit
\begin{equation}
\beta\sigma_b(0)\;=\,-i\,\Delta\rho\; \frac{C_b(0)\,C_b'(i\kappa_b)}{\,2\;|C_b^{(3)}(0;i\kappa_b,0)|\,}
\label{sigma0}
\end{equation}
where $C_b'(q)=dC_b(q)/dq$. This result follows from expanding $C_b(q_*)$ about the isolated pole at $i\kappa_b$:
\begin{equation}
C_b(q_*)\;=\;-\frac{i\,q^2}{2\kappa_b}\;C_b'(i\kappa_b)\;+\;\mathcal{O}(q^4)
\label{Cexpand}
\end{equation}
From the form of the result (\ref{sigma0}), we learn that the values of $\sigma_l(0)$ and $\sigma_g(0)$ are not necessarily equal to the equilibrium tension $\sigma$ (although in certain approximations this will turn out to be the case). The reason for this is that the {\it{far}} behaviours (\ref{Slargel}) and (\ref{Slargeg}) refer to the limit $|z|\to\infty$ at {\it{fixed}} $q>0$, while the Wertheim result (\ref{SGM}) (setting $g=0^+$), refers to $q\to 0$ at {\it{fixed}} $z$. The {\it{cross-over}} between these distinct limits can be understood by considering the  leading-order correction to the far behaviours (\ref{Slargel}) and (\ref{Slargeg}) occurring at small $q$. Expressions (\ref{Slargel}) and (\ref{Slargeg}) are  particular solutions of the Ornstein-Zernike equation, and one may add to them any solution for which the LHS of (\ref{OZS}) vanishes, i.e. the complementary function. At $q=0$, the Yvon equation (\ref{YvonC}), identifies the complementary function as being proportional to $\rho'(z)$ which therefore displays the same exponential decay appearing in (\ref{Slargel}) and (\ref{Slargeg}). However, at {\it{non-zero}} $q$, the complementary function must decay faster than $\exp({-\kappa_b |z|})$. Employing the same density expansion around the bulk value of the direct correlation function, and using a similar Bromwich integral trick, it is straightforward to see that, far from the interface, the complementary function must behave as $\exp({-\sqrt{\kappa_b^2+q^2} |z|})$. Thus, including the leading and next-to-leading order exponential decays in $z$, the local structure factor behaves as\vspace*{.15cm}
\begin{equation}
\begin{split}
& \hspace{2.cm} S(z;q)\;=\; S_b(q)+\\[.15cm] &\,\frac{\rho'(z)\,\Delta \rho}{\;\beta\, q^2\,}\left(\frac{1}{\sigma_b(q)}+\left[\frac{1}{\sigma}-\frac{1}{\sigma_b(0)}+\mathcal{O}(q^2)\right]e^{-\Delta\kappa_b(q)|z|}\right)\,+\,\cdots
\label{Scrossover}
\end{split}
\end{equation}
where $\Delta \kappa_b(q)\equiv \sqrt{\kappa_b^2+q^2}-\kappa_b \approx q^2/2\kappa_b$. This expression describes the cross-over from the far behaviour, characterised by the wavevector dependent tensions $\sigma_b(q)$, to the Wertheim-like Goldstone mode divergence as $q\to 0$, involving the equilibrium tension $\sigma$. Note that the terms written as order $\mathcal{O}(q^2)$ in (\ref{Scrossover}) are irrelevant when $z$ is greater than a few bulk correlation lengths from the interface, due to the exponential factor $\exp({-\Delta\kappa_b(q)|z|})$.\\

\subsection{Remarks and repercussions}

{\bf a)} The above results show that the wavevector dependence of $\sigma_l(q)$ and $\sigma_g(q)$ is associated with two-body and three-body correlation functions in the {\it{bulk}} liquid and gas phases. In order to proceed further, we next draw on mean-field DFT and consider a standard Helmholtz free-energy functional written as \cite{Evans1990}
\begin{equation}
\mathcal{F}[\rho]\,=\,\mathcal{F}_h[\rho]\;+\;\frac{1}{2}\int\!\!\!\int\! d{\bf{r}}_1d{\bf{r}}_2\;\rho ({\bf{r}}_1)\, w(|{\bf{r}}_1-{\bf{r}}_2|)\, \rho ({\bf{r}}_2)
\label{DenFun}
\end{equation}
where $F_h[\rho]$ is the functional for hard-spheres which accounts for repulsive intermolecular forces. The second term accounts for attractive contributions, where $w(r)$ is the attractive part of the intermolecular potential. Since the attractive term is quadratic in the density $\rho({\bf{r}})$ it makes no contribution to $C_b^{(3)}$ and for the functional (\ref{DenFun}), the wavevector dependence of $\sigma_b(q)$ reduces to
\begin{equation}
q^2\,\sigma_b(q)\,\;\propto\; \frac{C_b(q)\, C_b(q_*)}{\;C_h^{(3)}(q;i\kappa_b,0)\,}
\label{full}
\end{equation}
where $C_h^{(3)}$ is the three-body direct correlation function of the {\it{bulk}} hard-sphere fluid (evaluated at the appropriate liquid or gas density). At wavelengths much larger than the distance set by the hard-sphere diameter $d$, specifically $q\ll 2\pi/d$, one may reliably approximate the denominator as $C_h^{(3)}(0;i\kappa_b,0)$, which eliminates the  wavevector dependence from this term. Indeed, this simplification is realised explicitly within a local density approximation, where the repulsive contribution is approximated $\mathcal{F}_h[\rho]=\int\! d{\bf{r}}\,f_h\!\left(\rho({\bf{r}})\right)$, where $f_h(\rho)$ is the free-energy density for bulk hard-spheres. In this case, the three-body bulk correlation function is trivially determined as $C^{(3)}_b({\bf{r}}_1,{\bf{r}}_2,{\bf{r}}_3)=\beta\mu_h''(\rho_b)\,\delta({\bf{r}_{12}})\, \delta({\bf{r}_{13}})$ evaluated at $\rho_l$ or $\rho_g$, so that the double Fourier transform is wavevector independent. Here $\mu_h(\rho_b)\equiv f_h'(\rho_b)$ is the chemical potential for bulk hard-spheres and the prime denotes the derivative w.r.t density $\rho$. For larger wavevectors $q\to \pi/d$ one must improve upon the local density approximation for $F_h[\rho]$. This has been discussed, within the context of non-local DFT for hard-spheres, by several authors - see Sec.~IV of \cite{EvansDFT1992}. We do not consider such improvements further since we do not believe they are of concern to the most important properties of $\sigma_b(q)$.\\

Using the local density approximation for the three-body term $C_h^{(3)}$, we arrive at our final (approximate) results which we can summarise as follows: On each side of the interface $S(z;q)$ decays exponentially as $|z|\to\infty$,
\begin{equation}
S(z;q)\;=\; S_b(q)\;+\;\frac{\rho'(z)\,\Delta \rho}{\;\beta\,\sigma_b(q)\, q^2\,}\;+\;\cdots
\label{Slargegen}
\end{equation}
with distinct liquid and gas wavevector dependent tensions identified as
\begin{equation}
q^2\,\sigma_b(q)\;\propto\; C_b(q)\, C_b(q_*),
\label{LDA0}
\end{equation}
or, equivalently,
\begin{equation}
\frac{1}{\,q^2\,\sigma_b(q)\,}\;\propto\; S_b(q)\, S_b(q_*)
\label{LDA}
\end{equation}
Similarly the expression (\ref{sigma0}) for the $q\to 0$ limit of $\sigma_b(q)$ reduces to
\begin{equation}
\beta\sigma_b(0)\;=\,-i\,\Delta\rho\; \frac{C_b(0)\,C_b'(i\kappa_b)}{2\beta\;|\mu_h ''(\rho_b)|\,}
\label{sigma0LDA}
\end{equation}
which, we emphasize, involves only bulk quantities. These are the central results of our paper and are the basis for all further interpretation. Provided that $q\ll\pi/d$, (\ref{LDA0}), or equivalently (\ref{LDA}), should yield an excellent approximation to the formal results (\ref{Gen}) and (\ref{Geng}), and involve only easily accessible two-body terms. \\

{\bf b)} Perhaps the most important implication of the above analysis is that, far from the interface, the local structure factor separates {\it{exactly and unambiguously}} into a bulk contribution and an interfacial contribution. Eqns.~(\ref{Slargel}) and (\ref{Slargeg}) display the same position dependence as the translational eigenfunction $\rho'(z)$. Thus the interfacial contribution, beyond mean-field, shows the Capillary-Wave scaling behaviour (\ref{Scaling}). We emphasize that the unambiguous separation of $S(z;q)$ into bulk and interfacial contributions {\it{far}} from the interface does not mean that $S(z;q)$ separates exactly at all positions $z$. This is clear from Eqn.~(\ref{Scrossover}) which includes the next-to-leading order exponential decay. It is only when this contribution is negligible that $S(z;q)$ separates. However with this caveat, at large distances, it is difficult to imagine how the separation of $S(z;q)$ in (\ref{Slargegen}) could be more clear cut. This is the key to all further analysis. It is evident that the price paid for this separation is two-fold. Firstly, one must accept that the wavevector dependent coefficients $\sigma_b(q)$, characterising the interfacial terms are {\it{different}} on the liquid ($b=l$) and gas ($b=g$) sides. Thus, one is forced to conclude that the expansions 
\begin{equation}
\begin{split}
&\sigma_l(q)\;=\;\sigma_l(0)\;+\;K_l\,q^2\;+\;\cdots\\[.1cm]
&\sigma_g(q)\;=\;\sigma_g(0)\;+\;K_g\,q^2\;+\;\cdots
\end{split}
\end{equation}
define \textit{distinct} liquid and gas rigidity coefficients. We emphasise that $\sigma_l(q)$ and $\sigma_g(q)$ are determined by the bulk properties of the respective liquid and gas phases. This means that, like the equilibrium tension $\sigma$ itself, the wavevector dependent tensions are not altered, or renormalized, by interfacial fluctuations. The asymptotic decay of $S(z;q)$ \textit{is} altered by interfacial fluctuations, but only through the broadening of the density profile, specifically its derivative $\rho'(z)$ entering (\ref{Slargegen}) which simply renormalizes the amplitudes $\alpha_l$ and $\alpha_g$, as mentioned above.\\
 
The second point that we are forced to accept is that, in the long wavelength limit, the values of the coefficients $\sigma_b(0)$, given by (\ref{sigma0LDA}), are {\it{not}} identically equal to the equilibrium surface tension. As we have emphasized earlier, this does not contradict the Wertheim prediction that $S(z;q)\to \rho' (z)\Delta\rho/\beta\sigma q^2$ as $q\to 0$ since there is a simple cross-over to this asymptotic divergence involving the next-to-leading order exponential term, as shown in (\ref{Scrossover}). However, it is only the {\it{leading order}} exponential decay of $S(z,q)$ which gives a clear and unambiguously separation into bulk and interfacial contributions, allowing us to define and identify the wavevector dependent tensions $\sigma_b(q)$. The relation between $\sigma_b(0)$ and the equilibrium tension $\sigma$ is discussed at length when we describe our model calculations below.\\ 
 
 {\bf c)} For small $q$ one can the expand the term $C_b(q_*)$, appearing in eqn. (\ref{LDA0}) for $\sigma_b(q)$, about the isolated pole at $i\kappa_b$ as in (\ref{Cexpand}). The higher-order terms constitute one source of the wavevector dependence of $\sigma_b(q)$. However, it also follows from (\ref{LDA0}) that at temperatures close to $T_c$, where the bulk correlation length $\xi_b$ is large, the dominant contribution to $\sigma_b(q)$ must come from the $q$ dependence of $C_b(q)=1/S_b(q)\propto 1+\xi_b^2 q^2+ \mathcal{O}(q^4)$. In addition, our model calculations indicate that at these temperatures $\sigma_b(0)\approx\sigma$. This means that the next-to-leading order exponential term shown in (\ref{Scrossover}) is unimportant, so that at high temperatures 
\begin{equation}
\sigma_b(q)\;\approx\; \sigma\; +\;\sigma\, \xi_b^2\, q^2\;+\;\cdots
\end{equation}
Moreover, as we shall see below, within square-gradient theory with an explicit double quartic free-energy density, $\sigma_b(q)=\sigma (1+\xi_b^2 q^2) $ is the full result for all temperatures. Hence, close to $ T_c$, the liquid and gas rigidities must behave as
\begin{equation}
K_l\,\approx\, \sigma\,\xi_l^2,\hspace{1cm} K_g\,\approx\, \sigma\,\xi_g^2
\label{Ksapprox}
\end{equation}
both of which are obviously positive. Although our present analysis is based on a mean-field treatment, we might expect a result similar to (\ref{Ksapprox}) to hold beyond mean-field. Invoking hyperscaling \cite{Evans1992,Das2011}, we thus expect that the rigidity coefficients approach a finite, positive value , as $T\to T_c^-$. The simulation results of H\"ofling and Dietrich \cite{Hofling2015} shown in Fig.~\ref{Fig1} are consistent with this scaling prediction. However, our prediction is completely at odds with the theory of Refs.~\cite{Blokhuis2008,Blokhuis2009} which predicts a negative $K$ that vanishes in the same fashion as the tension $\sigma$ as $T\to T_c^-$. We return to the critical value of $K_b$ at the end of our paper. \\

{\bf d)} It is not possible to reconcile these results from microscopic theory with the central assumption of extended capillary-wave models, namely that the fluctuations of the interface can be understood using a single wavevector dependent surface tension. Moreover, the positive values of the rigidities, defined via $S(z;q)$, highlight the inadequacy of any attempt to define a rigidity using extended effective Hamiltonian theory as derived from DFT. This is the case even when $K_l=K_g$, as for perfect Ising or lattice-gas symmetry. To see this, we recall that within a DFT framework an interfacial Hamiltonian is identified as $H[\ell]=\Omega_V[\rho_\Xi]-\Omega_V[\rho(z)]$ where $\rho(z)$ is the equilibrium  density profile and $\rho_\Xi$ is the profile that minimizes the grand potential subject to the constraint on the non-planar interface location \cite{Fisher1991,Parry1994}. This constrained profile is a function of $\bf r$, a functional of $\ell({\bf x})$, and also depends on the precise definition of the interface location. However, we know that $\rho_\Xi\approx\rho\big(z\!-\!\ell({\bf{x}})\big)$ for long wavelength fluctuations, and substitution into $\Omega_V[\rho_\Xi]$ identifies \cite{Parry1994} the parameter $\sigma$ in (\ref{CWreal}) as the exact Triezenberg-Zwanzig expression (\ref{TZ}) for the equilibrium surface tension. Thus, at long wavelengths, any $H[\ell]$ derived from DFT is fully consistent with the original macroscopic Capillary-Wave Hamiltonian (\ref{CWreal}). However, the same ansatz for $\rho_\Xi$ now generates an upper bound \cite{Parry1994} for the rigidity
\begin{equation}
\beta\mathcal{K}_\textsc{\tiny ECW}^\textsc{\tiny DFT}\;<\; \int\!\!\!\int\! dz_1\, dz_2\;\,\rho'(z_1)\, C_4(z,z')\,\rho'(z_2)
\label{Kecw}
\end{equation}
which involves the the fourth moment of the direct correlation function. Here, we have added the superscript to emphasise that this is the DFT prediction for the rigidity appearing in an extended capillary-wave effective Hamiltonian. Within simple square-gradient theories, $C_4(z,z')$ is zero implying that $\mathcal{K}_\textsc{\tiny ECW}^\textsc{\tiny DFT}< 0$ for {\it{any}} definition of the interface position. The same is also true for the Sullivan local density functional model, for which $C_{4}(z,z')$ is negative (see later). In other words, it appears that the effective Hamiltonian rigidity $\mathcal{K}_\textsc{\tiny ECW}^\textsc{\tiny DFT}$ is not related to the rigidities $K_l$ and $K_g$ defined through the large distance behaviour of the local structure factor.\\

\section{Square-gradient theory and the Sullivan model.}

In this section, we determine $\sigma_l(q)$ and $\sigma_g(q)$ within square-gradient theory and also the Sullivan (local density functional) model. In both cases, the asymptotic decay of $S(z;q)$ can be determined from explicit solution of the appropriate Ornstein-Zernike equation. Throughout this section, we set $k_B T=1$ for convenience.\\

\subsection{Square gradient theory.}
 Within square-gradient theory, the Grand potential functional (\ref{Omegagen}) is given by 
\begin{equation}
\Omega_V[\rho]=\int\!\!d{\bf r}\; \left\{\,\frac{f_2}{2} (\nabla \rho)^ 2 + \phi(\rho) \right\}
\label{LGW}
\end{equation}
In general the coefficient of the gradient term, $f_2$, is proportional to the second moment of the bulk direct correlation function and, in principle, depends on the density \cite{Evans1979}. However, this dependence is weak, and $f_2$ can be reliably approximated by a constant, at least away from the bulk critical region. For example, if one begins with the mean-field functional (\ref{DenFun}), makes the local density approximation to $F_h[\rho]$, it follows that $f_2=-\frac{2\pi}{3}\int \!dr\, r^4 w(r)$ which is a constant \cite{Lu1985}. The free-energy density $\phi(\rho)$ is a double-well potential that models the coexistence of phases, while the curvatures $\phi '' (\rho_g)/f_2=\kappa_g^2$ and $\phi '' (\rho_l)/f_2=\kappa_\ell^2$ identify the inverse bulk correlation lengths of the gas ($\kappa_g\equiv 1/\xi_g$) and liquid ($\kappa_l\equiv 1/\xi_l$) phases, respectively.
Similarly the bulk structure factor $S_b(k)$ in the liquid ($b=l$) and gas ($b=g$) phases is given by
\begin{equation}
S_b(k)\;=\;\frac{S_b(0)}{\;1+k^2\,\xi_b^2\,}
\label{SGTS}
\end{equation}
where $S_b(0)=1/\phi''(\rho_b)$. Fourier transforming, we see that the second-moment and true correlation lengths are identical in this theory.
The Euler-Lagrange equation for the profile is obtained by minimising (\ref{LGW})
\begin{equation}
f_2\,\frac{\,d^2\rho(z)}{dz^2}\; =\; \phi' (\rho)
\label{EL}
\end{equation}
which is solved subject to boundary conditions $\rho(-\infty)=\rho_g$ and $\rho(\infty)=\rho_l$. From (\ref{EL}), it is easy to see that $\rho(z)$ decays exponentially on either side of the interface, as in (\ref{rhodecayl}) and (\ref{rhodecayg}). The Euler-Lagrange equation has a familiar first integral and determines the tension as
\begin{equation}
\sigma\;=\;f_2\int\! dz\; \rho'(z)^ 2
\label{SGTsigma}
\end{equation}
which is equivalent to the well known expression \cite{Rowlinson1982}
\begin{equation}\label{SGTsigma2}
\sigma\;=\;\int\! d\rho\; \sqrt{2\,f_2 \big(\phi(\rho)-\phi(\rho_g)\big)}
\end{equation}
where the integral is from $\rho_g$ to $\rho_l$. The Fourier transform of the direct correlation function follows using (\ref{Cdef}) and (\ref{CFourierdef})
\begin{equation}
C(z,z';q)\;=\;\big(-f_2\,\partial^2_z+f_2\,q^ 2 +\phi''(\rho(z))\big)\,\delta(z-z')
\label{CSGT}
\end{equation}
so that the Ornstein-Zernike equations (\ref{OZeqn}) and (\ref{OZS}) for $G(z,z';q)$ and $S(z;q)$ are
\begin{equation}
\big(-f_2\,\partial^2_z+f_2\,q^ 2 +\phi''(\rho(z))\big)\,G(z,z';q)\;=\;\delta(z-z')
\label{OZ}
\end{equation}
and
\begin{equation}
\big(-f_2\,\partial^2_z+f_2\,q^ 2 +\phi''(\rho(z))\big)\,S(z;q)\;=\;1
\label{OZSSGT}
\end{equation}
respectively. The solution for $G(z;z';q)$ can be written as the spectral expansion
\begin{equation}
G(z,z';q)\;=\;\sum_{n=0} \frac{\,\Psi_n^*(z)\Psi_n(z')}{E_n+f_2\,q^2}
\label{spectral}
\end{equation}
where the normalised eigenfunctions satisfy
\begin{equation}
\left(-f_2\,\partial^2_{z}+\phi''(\rho(z)\right)\,\Psi_n(z)\,=\,E_n\, \Psi_n(z)
\end{equation}
Inspection of the profile equation (\ref{EL}) identifies the ground state  eigenfunction $\Psi_0(z)=c\,\rho'(z)$, which has energy $E_0=0$, in addition to higher energy bound and scattering states \cite{Evans1981}. Normalization of $\Psi_0$ and (\ref{SGTsigma}) determines $c^2=f_2/\sigma$. Hence, we can be sure that for fixed $z$ and $q\to 0$, the $n=0$ term guarantees 
\begin{equation}
S(z;q)\;=\;\frac{\,\rho'(z)\,\Delta \rho\;}{\sigma q^2}\;+\;\mathcal{O}(q^0)
\label{spectral2}
\end{equation}
in accordance with Wertheim's exact sum-rule expectation (\ref{SGM}). It is also illuminating to note from (\ref{CSGT}) that the second moment $C_2(z,z)=f_2\delta (z-z')$ which, on substituting into (\ref{TZ}), recovers (\ref{SGTsigma}) for the tension $\sigma$, attesting to the overall self consistency of the square gradient theory \cite{Evans1979}.\\

In order to obtain the full wavevector dependence for $z$ far from the interface, we solve the second-order differential equation (\ref{OZSSGT}) by writing, in standard fashion, $S(z;q)=S^\textsc{\tiny CF}(z;q)+S^\textsc{\tiny PS}(z;q)$. For example, on the liquid side of the interface ($z\to\infty$), we substitute the asymptotic decay (\ref{rhodecayl}) of the profile $\rho(z)$ into (\ref{OZSSGT}) and expand
\begin{equation}\label{expand}
\phi''(\rho(z))\;=\; f_2\, \kappa_l^2\;-\;\alpha_l\,\phi'''(\rho_l)\,e^{-\kappa_l z}\;+\;\cdots
\end{equation}
where the higher-order terms are $\mathcal{O}(e^{-2\kappa_l z})$. Elementary techniques, for example using a WKB approximation, determine that the complementary function $S^\textsc{\tiny CF}(z;q)$ decays as $e^{-\sqrt{\kappa_l^2+q^2}\, z}(1+\mathcal{O}(e^{-\kappa_l\,z}))$ and can be neglected in comparison with the particular solution. Note that this is precisely in accord with the general prediction (\ref{Scrossover}) which includes the next-to-leading order correction. The particular solution can be obtained from the ansatz 
\begin{equation}
S^\textsc{\tiny PS}(z;q)\;=\;S_l(q)\,+\,a_l(q)\,e^{-\kappa_l z}\;+\;\mathcal{O}(e^{-2\kappa_l z})
\label{2}
\end{equation}
where the coefficient $a_l(q)$ is determined from simple substitution of (\ref{expand}) and (\ref{2}) into (\ref{OZSSGT}),
\begin{equation}
a_l(q)\;=\;\frac{\;\alpha_l\,\phi'''(\rho_l)\,S_l(q)\,}{f_2\,q^2}
\end{equation}
and we have used (\ref{SGTS}). Then, after recasting the position dependence of (\ref{2}) in terms of the derivative of the density profile, $\rho' (z)\approx\kappa_l\,\alpha_l\, e^{-\kappa_l\,z}$, we find that the asymptotic decay of $S(z;q)$ as $z\to\infty$ is
\begin{equation}
S(z;q)\;=\; S_l(q)\;+\;\frac{\rho'(z)\,\Delta \rho}{\;\sigma_l(0)\, q^2(1+\xi_l^2q^2)\,}\;+\;\cdots
\label{Slargelsgt}
\end{equation}
with, the coefficient $\sigma_l(0)$ determined as
\begin{equation}
\sigma_l(0)=\frac{f_2\,\Delta\rho\, \kappa_l\, \phi''(\rho_l)}{\phi '''(\rho_l)}
\label{3}
\end{equation}
The higher order terms in (\ref{Slargelsgt}) decay as $\mathcal{O}(e^{-2\kappa_l z},e^{-\sqrt{\kappa_l^2+q^2}\, z})$.
Of course, a similar result applies deep in the gas region, for which, as $z\to-\infty$,
\begin{equation}
S(z;q)\;=\; S_g(q)\;+\;\frac{\rho'(z)\,\Delta \rho}{\;\sigma_g(0)\, q^2(1+\xi_g^2q^2)\,}\;+\;\cdots
\label{Slargegsgt}
\end{equation}
with 
\begin{equation}
\sigma_g(0)\;=\; - \frac{f_2\,\Delta\rho\, \kappa_g\, \phi''(\rho_g)}{\phi '''(\rho_g)}
\label{4}
\end{equation}
Comparison with the definitions (\ref{Slargel}) and (\ref{Slargeg}) identifies the liquid and gas wavevector dependent tensions as
\begin{equation}
\sigma_l(q)=\sigma_l(0)(1+q^2\xi_l^2)\hspace{0.25cm}\textup{and}\hspace{0.25cm}\sigma_g(q)=\sigma_g(0)(1+q^2\xi_g^2)
\label{5}
\end{equation}
both of which show a simple monotonic increase with $q$. Introducing, within square gradient theory, $\mu(\rho_b)\equiv \phi'(\rho_b)$, it follows that $\partial\rho_b/\partial\mu=\phi''(\rho_b)^{-1}$, proportional to the compressibility, and hence that $\partial^2\rho_b/\partial\mu^2=-\phi'''(\rho_b)/\phi''(\rho_b)^3$. Thus, the results (\ref{3}) and (\ref{4}) for $\sigma_l(0)$ and $\sigma_g(0)$ can be re-expressed simply 
in terms of bulk thermodynamic quantities as
\begin{equation}
\sigma_b(0)\;=\, \xi_b\,\Delta\rho\;\frac{\partial\rho_b}{\partial\mu}\big/\left|\frac{\partial^2\rho_b}{\partial\mu^2}\right|
\label{sigma0sgt}
\end{equation}
where the derivatives are performed at constant $T$.
The wavevector dependence of $\sigma_l(q)$ and $\sigma_g(q)$ shown in (\ref{5}) is in precise agreement with the general prediction (\ref{LDA}) since simple substitution for $q_*$ gives 
\begin{equation}
S_b(q)\,S_b(q_*)\;\propto\;\frac{1}{\;q^2(1+\xi_b^2q^2)\,}
\end{equation}
Notice that, in this case, $S_b(q_*)\propto q^{-2}$ generates a pure Goldstone mode divergence. Therefore, the {\it{entire}} wavevector dependence of $\sigma_b(q)$ arises from that of $C_b(q)\propto (1+q^2\xi_b^2)$. Similarly, the results (\ref{3}) and (\ref{4}), or equivalently (\ref{sigma0sgt}), for $\sigma_b(0)$ are in precise agreement with the general expression (\ref{sigma0}). It is evident that for a {\it{general}} double well potential $\phi(\rho)$ the results (\ref{3}) and (\ref{4}) for $\sigma_l(0)$ and $\sigma_g(0)$ are not identical to the expression (\ref{SGTsigma2}) for the equilibrium surface tension $\sigma$. However, a remarkable aspect of the present analysis is that, for the standard Landau free energy density  
\begin{equation}
 \phi(\rho)=-\frac{t}{2}\,(\rho-\rho_c)^2\,+\,\frac{u}{4}\,(\rho-\rho_c)^4
\label{Landau}
\end{equation}
where $\rho_c$ is the critical density, $t\propto T_c-T$ and $u$ a positive parameter, the integral (\ref{SGTsigma2}) determining $\sigma$ and the expression (\ref{3}), or equivalently (\ref{sigma0sgt}), for $\sigma_l(0)$ (which is equal to $\sigma_g(0)$), yield the same value $B_\textsc{\tiny FW}\,f_2\,\kappa_b\,(\Delta\rho)^2$ where $B_\textsc{\tiny FW}\,=1/6$ is the mean-field value of the universal Fisk-Widom critical amplitude \cite{Rowlinson1982,Fisk1969}. That is, for the free-energy density (\ref{Landau}) characteristic of standard mean-field Landau theory, we obtain
 \begin{equation}
\sigma_b(0)=\sigma
\label{6}
\end{equation}
This means that the large distance decay of $S(z,q)$ can be written more simply as 
\begin{equation}
S(z;q)\;=\; S_b(q)\;+\;\frac{\rho'(z)\,\Delta \rho}{\;\sigma\, q^2(1+\xi_b^2q^2)\,}\;+\;\cdots
\label{7}
\end{equation}
which now also contains the Wertheim limit as $q\to 0$, since the higher order terms in (\ref{7}) are now entirely negligible for all wavevectors. Thus, the wavevector dependent tension is simply $\sigma_b(q)=\sigma(1+ \xi_b^2q^2)$, implying that the rigidity $K_b$ is positive and given by 
\begin{equation}
K_b\,=\,\sigma\, \xi_b^2\
\label{Ksgt}
\end{equation}
in agreement with (\ref{Ksapprox}). Of course, the potential (\ref{Landau}) has an Ising (lattice-gas) symmetry so that $\xi_l=\xi_g=\xi_b$. For more general potentials $\phi(\rho)$, it may not be the case that $\sigma_b(0)=\sigma$ at all temperatures. However, on the basis of universality, it is natural to anticipate that for $T$ close to $T_c$ (\ref{Landau}) is increasingly realistic and that, at least at mean-field level, $\sigma_b(0)\approx \sigma$ on both the liquid and gas sides of the interface. We shall return to this conjecture in the concluding section of the paper.\\

The above results are quite general in that they apply to all square-gradient theories of the interfacial region, except when the third derivative $\phi'''(\rho_b)$ vanishes. Therefore, these results do not apply to the double parabola approximation for $\phi(\rho)$ introduced in  \cite{Parry2014}. Note however that the exact form of $S(z;q)$ is known for this approximate treatment \cite{Parry2014}; it decays as $\exp({-\sqrt{\kappa_b^2+q^2} |z|})$ which is consistent with (\ref{Scrossover}) since $\sigma_b(0)=\infty$.\\

\subsection{An exactly solvable potential within square gradient theory}

\begin{figure}[b]
\includegraphics[width=\columnwidth]{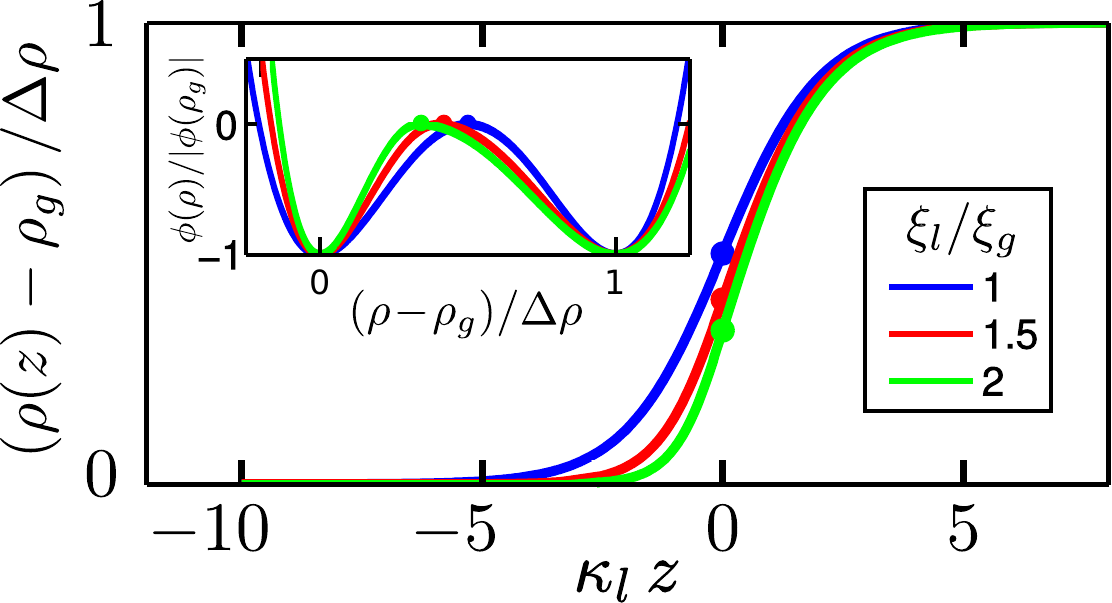}
\caption{\label{Fig2} Density profiles $\rho(z)$ obtained within the double-quartic square-gradient theory for different liquid-gas asymmetries as measured by the ratio of bulk correlation lengths $\xi_l/\xi_g$. Inset: The corresponding double-quartic potentials $\phi(\rho)$; the two minima lie at the coexisting densities $\rho_g$ and $\rho_l$. The circles represent the points where the density gradient is maximum, corresponding to $\rho(0)=\rho_0$.}
\end{figure}

\begin{figure*}[t]
\includegraphics[width=\textwidth]{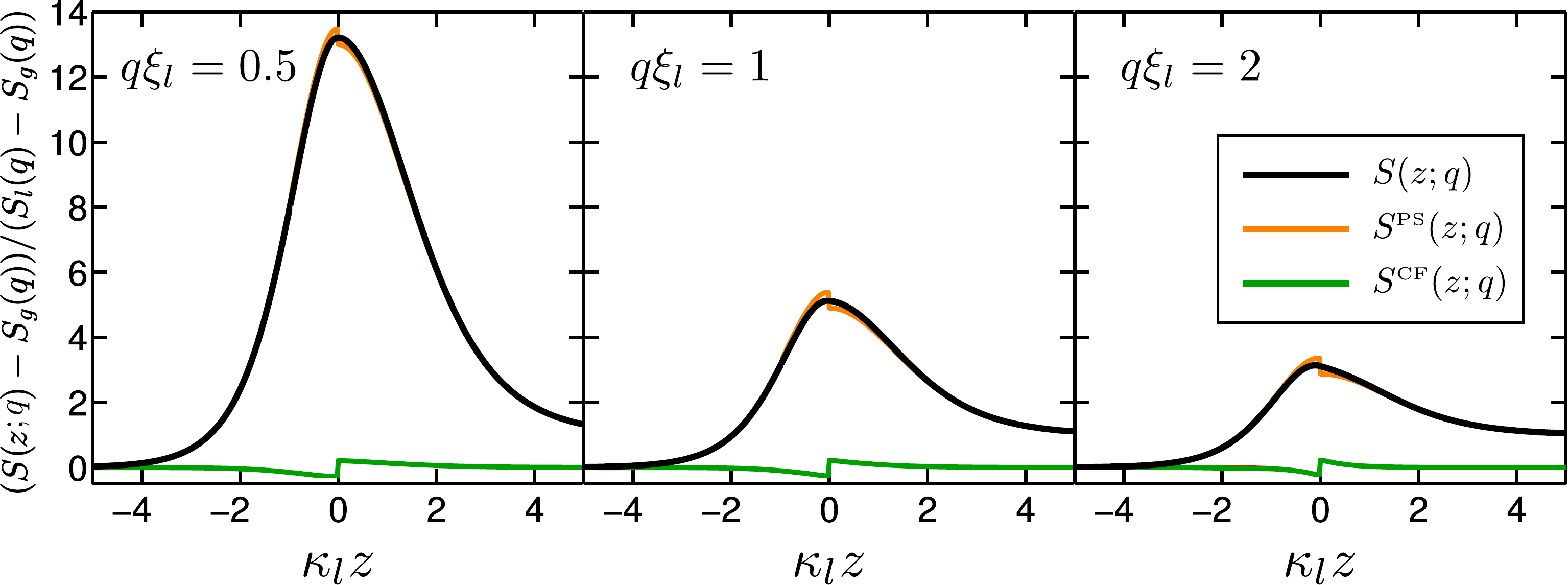}
\caption{\label{Fig3} The local structure factor $S(z;q)$ as a function of $z/\xi_l$ for wavevectors $q\xi_l=0.5$, $1$ and $2$, for the double-quartic square gradient theory with $\xi_l=1.5\xi_g$. The particular solution $S^\textup{\tiny PS}(z;q)$ is given by (\ref{SPS}) and accounts for the large distance behaviour, containing $\sigma_l(q)$ and $\sigma_g(q)$, and also the Wertheim-like Goldstone mode as $q\to 0$. Note that $S^\textup{\tiny PS}(z;q)$ well describes $S(z;q)$ even close to the interface. Also shown is the contribution from the complementary function $S^\textup{\tiny CF}(z;q)$ given by (\ref{SCFG}) and (\ref{SCFL}), which is always small. Although both $S^\textup{\tiny PS}(z;q)$ and $S^\textup{\tiny CF}(z;q)$ are discontinuous at $z=0$, the sum $S(z;q)$ is continuous and differentiable.}
\end{figure*}

Here we introduce a simple potential $\phi(\rho)$ entering (\ref{LGW}), for which $S(z;q)$ can be determined analytically. The model potential is defined using an asymmetric double-quartic 
\begin{equation}
\phi(\rho)=\left\{
\begin{array}{l}
 -\displaystyle\frac{t_g}{2}\,(\rho-\rho_0)^2\,+\,\frac{u_g}{4}\,(\rho-\rho_0)^4,  \hspace{0.7cm}\rho<\rho_0\\[.4cm]
-\displaystyle\frac{t_l}{2}\,(\rho-\rho_0)^2\,+\,\frac{u_l}{4}\,(\rho-\rho_0)^4, \hspace{0.7cm}\rho>\rho_0
\end{array}\right.
\label{SADQ}
\end{equation}
where $\rho_0$ is the value of the density  where the gradient $\rho'(z)$ is largest, which we place at $z=0$. For $t_g,t_l>0$ (and positive $u_g, u_l$), each branch of this double-quartic potential has a single minimum; thus, $\phi(\rho)$ has a minimum at the bulk gas density $\rho_g=\rho_0-\sqrt{t_g/u_g}$, with curvature $\phi''(\rho_g)=2t_g$ and value $\phi(\rho_g)=-t_g^2/4u_g$. Similarly, there is a minimum at the bulk liquid density $\rho_l=\rho_0+\sqrt{t_l/u_l}$, with curvature $\phi''(\rho_l)=2t_l$ and value $\phi(\rho_l)=-t_l^2/4u_l$. By setting $\phi(\rho_g)=\phi(\rho_l)$, which is equivalent to the condition
\begin{equation}
\rho_0\;=\;\frac{\kappa_g}{\kappa_g+\kappa_l}\,\rho_g\;+\;\frac{\kappa_l}{\kappa_g+\kappa_l}\,\rho_l,
\end{equation}
we ensure that the bulk gas and liquid phases coexist. The potential is smooth in the sense that both $\phi(\rho)$ and $\phi'(\rho)$ are continuous with both branches meeting at $\rho_0$, where $\phi(\rho_0)=\phi'(\rho_0)=0$.  We may now chose any value for $\xi_l/\xi_g$ to model the asymmetry between the bulk properties. The equilibrium density profile that follows from (\ref{EL}) is given by
\begin{equation}
\rho(z)=\left\{
\begin{array}{l}
\rho_0\,+\,(\rho_l\!-\!\rho_0)\,\tanh(\kappa_l z/2),\hspace{0.7cm}z>0\\[.4cm]
\rho_0\,-\,(\rho_g\!-\!\rho_0)\,\tanh(\kappa_g z/2),\hspace{0.7cm}z<0
\end{array}\right.
\end{equation}
which is continuous and differentiable at the origin (see Fig.~\ref{Fig2}). The surface tension in this model follows from (\ref{SGTsigma}) as
\begin{equation}
\sigma\;=\;\frac{f_2(\Delta\rho)^2}{\;3\,(\xi_g+\xi_l)\,}
\label{100}
\end{equation}
which vanishes as $(T_c-T)^{3/2}$ in accordance with standard mean-field expectations, including the correct critical amplitude \cite{Rowlinson1982}. The local structure factor can be determined directly from the Ornstein-Zernike equation (\ref{OZSSGT}) by writing it as sum of a particular solution and a complementary function
\begin{equation}
S(z;q)\;=\; S^\textsc{\tiny CF}(z;q)\;+\;S^\textsc{\tiny PS}(z;q)
\label{101}
\end{equation}
It is straightforward to show that, on each side of the interface, a particular solution is
\begin{equation}
S^\textsc{\tiny PS}(z;q)\;=\;S_b(q)\;+\;\frac{\rho'(z)\,\Delta \rho}{\;\sigma\, q^2(1+q^2\xi_b^2)\,}
\label{SPS}
\end{equation}
where $b=l$ for $z>0$, and $b=g$ for $z<0$ and that $\sigma$ is the equilibrium tension appearing in (\ref{100}). Standard techniques also determine that, on the gas side, the complementary function is
\begin{equation}\label{SCFG}
S_g^\textsc{\tiny CF}(z;q)=A_g(q) e^{\kappa_g(q)z}\Big(\tau_g(z)^2-\frac{2\kappa_g(q)}{\kappa_g}\tau_g(z)+1+\frac{4}{3}q^2\xi_g^2\Big)
\end{equation}
while, on the liquid-side,
\begin{equation}\label{SCFL}
S_l^\textsc{\tiny CF}(z;q)=A_l(q) e^{-\kappa_l(q)z}\Big(\tau_l(z)^2+\frac{2\kappa_l(q)}{\kappa_l}\tau_l(z)+1+\frac{4}{3}q^2\xi_l^2\Big)
\end{equation}
where we have abbreviated $\tau_b(z)=\tanh(\kappa_b z/2)$, and also
\begin{equation}
\kappa_b(q)\,=\,\sqrt{\kappa_b^2+q^2}
\end{equation}
The constants $A_b(q)$ follow from requiring that $S(z;q)$ given by the sum (\ref{101}) is continuous and differentiable at the origin, and simple matching of the solutions on the liquid and gas sides leads to
\begin{equation}
A_g(q)\;=\;\frac{\,S_g(q)-S_l(q)\,}{2 D_g(q)}
\label{Ag}
\end{equation}
where
\begin{equation}
D_g(q) =1+\frac{\xi_g}{\xi_l}\sqrt{\frac{1+q^2\xi_g^2}{1+q^2\xi_l^2}}+\frac{4}{3}q^2\xi_g^2\left(1+\frac{\xi_l}{\xi_g}\sqrt{\frac{1+q^2\xi_g^2}{1+q^2\xi_l^2}}\,\right)
\end{equation}
The expression for $A_l(q)$ is similar but has the subscripts $g$ and $l$ interchanged throughout.
The function $S(z;q)$ therefore changes smoothly between the bulk values $S_g(q)$ and $S_l(q)$ as $z$ moves through the interfacial region, and is maximal near the origin, where it displays the expected Goldstone mode singularity at small $q$. However, in general, the maximum in $S(z;q)$ does not occur exactly where $\rho'(z)$ is largest because of the wavevector dependence of the complementary function $S^\textsc{\tiny CF}(z;q)$.
The present calculation brings out an analogy with the familiar properties of a forced linear oscillator. The particular solution (\ref{SPS}), which separates neatly into bulk and interfacial contributions, is analogous to the steady-state behaviour, and determines the "far" properties of $S(z;q)$. Eqn.(\ref{SPS}) determines the separate liquid and gas wavevector dependent tensions as
\begin{equation}
\sigma_l(q)=\sigma(1+q^2\xi_l^2)\hspace{0.25cm}\textup{and}\hspace{0.25cm}\sigma_g(q)=\sigma(1+q^2\xi_g^2)
\label{102}
\end{equation}
Note that as for the Landau potential (\ref{Landau}), $\sigma_l(0)=\sigma_g(0)=\sigma$, but we no longer have Ising symmetry and the presence of different bulk correlation lengths on each side of the interface means that $\sigma_l(q)$ and $\sigma_g(q)$ are different. For the model potential (\ref{SADQ}) we therefore have different rigidities
\begin{equation}
K_l\,=\, \sigma\,\xi_l^2,\hspace{1cm} K_g\,=\, \sigma\,\xi_g^2
\label{Ksapprox2}
\end{equation}
in accordance with (\ref{Ksapprox}) based on general arguments. In turn, the complementary function, analogous to the "transient" behaviour, decays faster and is much smaller than the particular solution - see Fig.~\ref{Fig3}. This function displays a much more complicated wavevector dependence and $S^\textsc{\tiny CF}(z;q)$, given by (\ref{SCFG}) and (\ref{SCFL}), certainly does not separate into "bulk" and "interfacial" contributions unless we assign these arbitrarily. It is easy to show that, in the limit of $q\to 0$, the complementary function satisfies $S^\textsc{\tiny CF}(z;0)\propto \rho' (z)$ in agreement with our remarks made before (\ref{Scrossover}). However, the fact that, within the present model, $\sigma_b(0)=\sigma$ means that the complementary function does not contribute to Goldstone mode divergence as $q\to 0$. \\

\begin{figure}[t]
\includegraphics[width=\columnwidth]{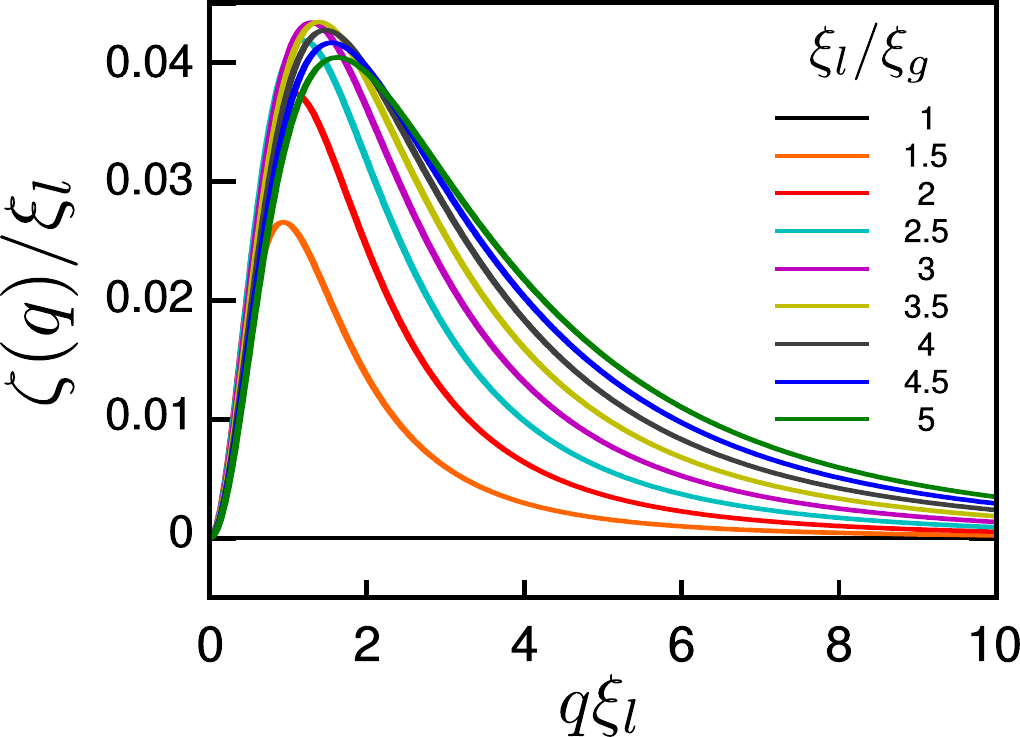}
\caption{\label{Fig4} The microscopic lengthscale $\zeta(q)$ weighting the difference in the bulk terms $S_l(q)$ and $S_g(q)$ in the background contribution to the total structure factor, obtained for the double-quartic potential, obtained for different liquid-gas asymmetries (see (\ref{StotQM}), (\ref{zetatrue})). This lengthscale is always small and displays a maximum when $q\approx\kappa_l$. The value of $\zeta(q)$ at this maximum depends on the asymmetry ratio $\xi_l/\xi_g$, and is largest $\approx 0.043\,\xi_l$ when $\xi_l/\xi_g\approx 3.25$.}
\end{figure}

In experimental and simulation studies focus is often placed on the total structure factor $S(q)$ obtained by integrating $S(z;q)$ over the macroscopic range $[-L_g,L_l]$ 
\begin{equation}
 S(q)=\int_{-L_g}^{L_l} \!\!\!dz\; S(z;q)
 \label{Stot0}
\end{equation}
Recall that for the present analysis the origin is located where the density gradient is largest. Within the asymmetric double-quartic model, $S(q)$ can be determined as
\begin{equation}\label{StotQM}
S(q)=S^{bulk}(q)+\frac{(\Delta\rho)^2}{ q^2} \left(\frac{f_g}{\sigma_g(q)}+\frac{f_l}{\sigma_l(q)}\right)-\zeta(q)\Delta S(q)
\end{equation}
where the wavevector dependent tensions are given by (\ref{102}), $S^{bulk}(q)=L_gS_g(q)+L_lS_l(q)$ is the bulk signal, $\Delta S(q)\equiv S_l(q)-S_g(q)$ and 
\begin{equation}
f_g=\frac{\rho_0-\rho_g}{\Delta\rho}\hspace{0.35cm}\textup{and} \hspace{0.35cm} f_l=\frac{\rho_l-\rho_0}{\Delta\rho}
\label{f}
\end{equation}
The first two terms in $S(q)$ arise from integration of $S^\textup{\tiny PS}(z;q)$. This gives the bulk contribution and a Goldstone mode term, the corrections to which come from the weighted combination of liquid and gas wavevector dependent surface tensions. The final term is generated by integration of $S^\textup{\tiny CF}(z;q)$ and, since $S^\textup{\tiny CF}(z;q)$ does not split into bulk and interfacial contributions, this must be deemed a "background" contribution. This is one of the reasons why defining or measuring a wavevector dependent tension through the total structure factor $S(q)$ is problematic \cite{Parry2015}. In the present case the microscopic lengthscale $\zeta(q)$ introduced in \cite{Parry2015} can be determined explicitly and after a straightforward but lengthy calculation we obtain
\begin{equation}
\zeta(q)=\frac{2+\frac{3}{2}q^2\xi_g^2}{\kappa_g(q)D_g(q)}-
\frac{2+\frac{3}{2}q^2\xi_l^2}{\kappa_l(q)D_l(q)}
\label{zetatrue}
\end{equation}
This displays a non-trivial wavevector dependence (see Fig.~\ref{Fig4}). Note that $\zeta(q)$ is always rather small and vanishes at large $q$, and also at small $q$ where it may be expanded $\;\zeta(q)\,\approx\,\frac{1}{2}\,q^2\xi_l\xi_g(\xi_l-\xi_g)$.\\

It is instructive to compare these results with those obtained for a square-gradient theory using a double parabola (DP) approximation for the potential $\phi(\rho)$ \cite{Parry2015}. The present work shows that the DP approximation is poor for the local structure factor, since it predicts, incorrectly, the asymptotic decay $S(z;q)-S_b(q)\propto e^{-\kappa_b(q)|z|}$  instead of $S(z;q)-S_b(q)\propto e^{-\kappa_b|z|}$, which is the more general result obtained here within square gradient theory. This means that the DP approximation misses the unambiguous separation of $S(z;q)$ occurring far from the interface expressed by (\ref{SPS}). Consequently, it overestimates the importance of the lengthscale $\zeta(q)$, which weights the bulk $S_b(q)$ in the background contribution to the total structure factor. For example, using the DP model we had  argued previously that $\zeta(0)$ is equal to the location of the Gibbs Dividing surface \cite{Parry2015}. It is now apparent that this is not true in general, and that the properties of $\zeta(q)$ are far {\it{less}} important than were indicated by the DP approximation. This is clear from Fig.~\ref{Fig4}, where the maximum value of $\zeta(q)$ within the present approximation is only $\zeta(q)\approx 0.043\xi_l$, and occurs when $\kappa_g/\kappa_l\approx 3.25$. This suggests that, in practice, the influence of $\zeta(q)$ on the background total structure factor is likely to be negligible. Certainly, the third term in (\ref{StotQM}) makes only a very small contribution in the present treatment.

\subsection{The Sullivan Model DFT}

Next we consider the Sullivan model described by the Helmholtz free-energy functional \cite{Sullivan1979,Sullivan1981,Tarazona1983}
\begin{equation}
\mathcal{F}[\rho]=\!\int\!\! d{\bf{r}}\,f_h\big(\rho({\bf{r}})\big)\,+\,\frac{1}{2}\int\!\!\!\int\! d{\bf{r}}_1d{\bf{r}}_2\;\rho ({\bf{r}}_1)\,
w(|{\bf{r}}_1-{\bf{r}}_2|)\, \rho ({\bf{r}}_2)
\label{300}
\end{equation}
where $f_h(\rho)$ is the free-energy density for bulk hard-spheres, and recall again that the  derivative $d f_h(\rho)/d\rho=\mu_h(\rho)$ is the local hard-sphere chemical potential, which is a monotonically increasing function of $\rho$. The first term in (\ref{300}) is the local density approximation to $F_h[\rho]$ in (\ref{DenFun}). In the Sullivan Model the intermolecular potential has a Yukawa form
\begin{equation}
w(r)\;=\;-\frac{\alpha}{\,4\pi rR^2\,}\;e^{-r/R}
\end{equation}
where $-\alpha\!=\!\int\! d{\bf{r}}\, w(r)$ is the integrated strength and the range is $R$. Hereafter, without loss of generality we set $R=1$ or, equivalently, measure all lengths in units of $R$. The Sullivan model does not describe properly the nature of short-ranged repulsion compared with non-local Fundamental Measure DFT models. However, such packing effects are much less pronounced for a free interface than, say, for adsorption near a wall. The density profile for a free interface follows from solving (\ref{Omegagen}) and (\ref{ELgen})
\begin{equation}
\mu\;=\;\mu_h\big(\rho({\bf{r}})\big)\;+\,\int\! d{\bf{r}}'\,\rho ({\bf{r}}')\; w(|{\bf{r}}-{\bf{r}}'|)
\end{equation}
which, on taking the Laplacian of both sides, reduces to
\begin{equation}
\nabla^2\mu_h\big(\rho({\bf{r}})\big)\;=\;\mu_h\big(\rho({\bf{r}})\big)\,-\,\mu\,-\,\alpha\,\rho({\bf{r}})
\end{equation}
and, thus, for a planar interface,
\begin{equation}\label{ELsull}
\frac{d^2\mu_h}{dz^2}\;=\,\mu_h\big(\rho(z)\big)\,-\,\mu\,-\,\alpha\,\rho(z)
\end{equation}
This is similar to the ODE for the profile in square-gradient theory (\ref{EL}). The surface tension $\sigma$, which can be written \cite{Sullivan1981}
\begin{equation}\label{SullTen}
\sigma\;=\;\frac{1}{\alpha}\int\! dz\, \left(\frac{d\mu_h}{dz}\right)^2,
\end{equation}
is similar to the square gradient result (\ref{SGTsigma}). We focus on the structure factor and note that the direct correlation function is given by
\begin{equation}
C({\bf{r}},{\bf{r}}')\;=\;\mu_h'\big(\rho({\bf{r}})\big)\,\delta({\bf{r}}-{\bf{r}}')\;+\;w(|{\bf{r}}-{\bf{r}}'|)
\label{Csull}
\end{equation}
where $\mu_h'(\rho)=d\mu_h/d\rho$. This has the 2D Fourier transform \cite{Tarazona1983,Parry1988}
\begin{equation}
C(z,z';q)\;=\;\mu_h'\big(\rho(z)\big)\,\delta(z-z')\,-\,\frac{\alpha\, e^{-\sqrt{1+q^2}\,|z-z'|}}{2\sqrt{1+q^2}}
\label{Cqsull}
\end{equation}
On substitution into the Ornstein-Zernike equation (\ref{OZS}), the integro-differential equation reduces, after taking two derivatives w.r.t $z$, to
\begin{equation}
\left(-\partial^2_z +q^2+1-\alpha \frac{d\rho}{d\mu_h}\right)
\Big(S(z;q)\,\mu_h'\big(\rho(z)\big)\Big)=\,1+q^2
\label{OZSsull}
\end{equation}
This simple ODE identifies the bulk structure factors $S_b(q)=1/C_b(q)$ in the liquid ($b=l$) and gas ($b=g$) phases as
\begin{equation}
S_b(q)=\frac{S_b(0)}{\;1+\displaystyle\frac{q^2\xi_b^2}{1+q^2R^2}\;} 
\label{310}
\end{equation}
where 
\begin{equation}
S_b(0)=\frac{1}{\;\mu_h'(\rho_b)-\alpha\;} 
\end{equation}
and identifies the Ornstein-Zernike correlation lengths $\xi_b$ as
\begin{equation}
\frac{R}{\xi_b}\;=\;\sqrt{\,\frac{\mu_h'(\rho_b)}{\alpha}-1\,}
\end{equation}
where we have reinstated the lengthscale $R$. The corresponding true correlation lengths are obtained using (\ref{pole}), with $\xi_b^T=\kappa_b^{-1}$. One finds that 
\begin{equation}
\kappa_b^2\, R^2\;=\;1-\frac{\alpha}{\mu_h'(\rho_b)}
\label{NEW}
\end{equation}
 which immediately gives $(\xi_b^T)^2 =\xi_b^2+R^2$ \cite{Lu1985}. Note that, at low temperatures, near the triple point, $\xi_b^T/\xi_b$ can be $\gg 1$, especially in the gas phase. To solve the second-order differential equation for $S(z;q)$ we define
 \begin{equation}
 H(z;q)\equiv \frac{S(z;q)\,\mu_h'\big(\rho(z)\big)}{1+q^2}
 \label{H}
\end{equation}
so that (\ref{OZSsull}) can be re-written 
\begin{equation}
\left(-\partial^2_z +q^2+1-\alpha \frac{d\rho}{d\mu_h}\right)
H(z;q)\;=\,1
\label{OZSsull2}
\end{equation}
This brings out clearly the analogy with equation (\ref{OZSSGT}) for the local structure factor in square gradient theory. Note that $d\rho/d\mu_h=(\mu_h '(\rho(z)))^{-1}$ is evaluated at the local density $\rho(z)$. One can determine $H(z,q)$ using
\begin{equation}
 H(z;q)=\int\! d z'\; H^{(2)}(z,z';q)
 \label{H2}
\end{equation}
where $H^{(2)}(z,z';q)$ is the Greens' function defined as
\begin{equation}
\left(-\partial^2_z +q^2+1-\alpha \frac{d\rho}{d\mu_h}\right)
H^{(2)}(z,z';q)\;=\,\delta(z-z') 
\label{Greensull}
\end{equation}
This has a spectral expansion, analogous to (\ref{spectral})
\begin{equation}
H^{(2)}(z,z';q)\;=\;\sum_{n=0} \frac{\,\Psi_n^*(z)\Psi_n(z')}{E_n+\,q^2}
\label{spectralsull}
\end{equation}
where the normalised eigenfunctions satisfy
\begin{equation}
\left(-\partial^2_z + 1-\alpha \frac{d\rho}{d\mu_h}\right)\,\Psi_n(z)\,=\,E_n\, \Psi_n(z)
\end{equation}

\begin{figure}[t]
\includegraphics[width=\columnwidth]{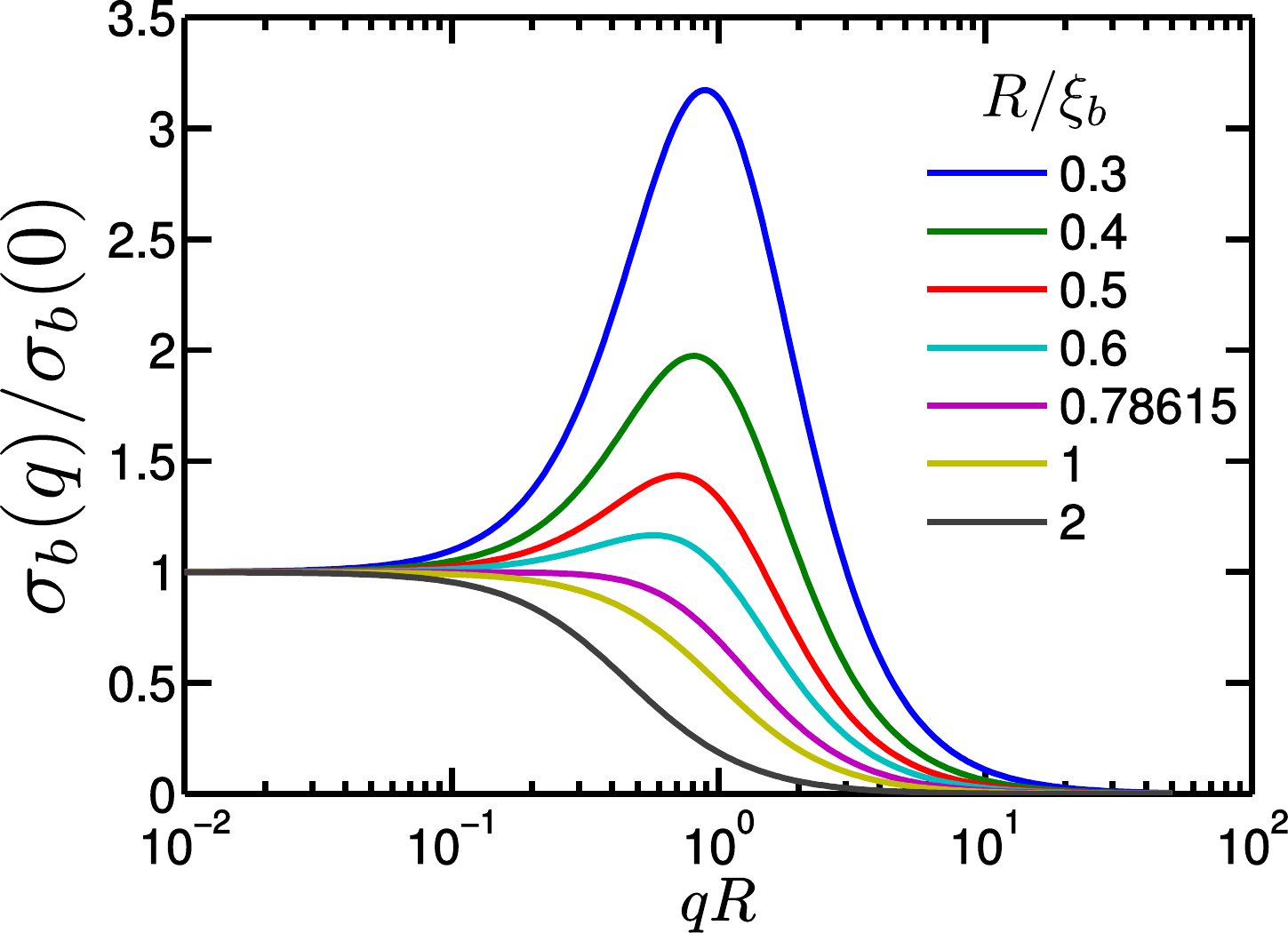}
\caption{\label{Fig5} Sullivan model prediction (\ref{sigmaqsull}) for the wavevector dependent tensions $\sigma_b(q)/\sigma_b(0)$ characterising the decay of $S(z;q)$ on the liquid ($b=l$) and gas ($b=g$) sides for different choices of $R/\xi_b$. Here, $\xi_b$ is Ornstein-Zernike bulk correlation length, and $R$ the (fixed) range of the potential. Thus $T/T_c$ is increasing from bottom to top curves. The rigidities $K_b$ are positive for $R/\xi_b<\sqrt{(\sqrt{5}-1)/2}\approx 0.78615$. }
\end{figure}

In this case, inspection of the profile equation (\ref{ELsull}) identifies the ground state  eigenfunction as $\Psi_0(z)=c\;d \mu_h/d z$, which has energy $E_0=0$. Here, the normalization constant follows from the Sullivan model result for the surface tension (\ref{SullTen}) as $c^2=1/\alpha\sigma$. As for square-gradient theory, the $n=0$ term dominates the spectral expansion as $q\to 0$ and, combining (\ref{H}), (\ref{H2}) and (\ref{spectralsull}) we obtain, in this limit,
\begin{equation}
S(z;q)\;=\;\frac{\,\rho'(z)\,\big(\mu_h(\rho_l)-\mu_h(\rho_g)\big)\;}{\alpha\sigma q^2}\;+\;\mathcal{O}(q^0)
\label{spectralsull2}
\end{equation}
Recalling that the liquid and gas densities satisfy $\mu_{co}=\mu_h(\rho_b)-\alpha\rho_b$, where $\mu_{co}$ is the chemical potential at bulk coexistence, one sees that equation (\ref{spectralsull2}) is identical to Wertheim's exact sum-rule prediction (\ref{SGM}) or (\ref{spectral2}).\\
To obtain the full wavevector dependence of $S(z;q)$ for $z$ far from the interface, we employ the same method used earlier for the square gradient theory. To this end, we continue to use the function $H(z;q)$. Consider, for example, the decay of $H(z;q)$ as $z\to\infty$. In this limit, one can expand the term $1-\alpha \frac{d\rho}{d\mu_h}$ appearing in (\ref{OZSsull2}) about the bulk liquid density to obtain
\begin{equation}
\left(-\partial^2_z +q^2+\kappa_l^2+\alpha\,\frac{\mu_h''(\rho_l)}{\mu_h '(\rho_l)^2}\,\delta\rho(z)+\cdots\right)
H(z;q)\;=\,1
\label{OZSsull3}
 \end{equation}
and substitute $\delta\rho\propto e^{-\kappa_l z}$ for the decay of the profile. The solution to this can be obtained from the ansatz, equivalent
to (\ref{2}),
 \begin{equation}
 H(z;q)\;=\;H_l(q)\;+\;\tilde a_l(q)\; e^{-\kappa_l z}+\cdots
 \label{Hansatz}
\end{equation}
where $H_l(q)=1/(\kappa_l ^2+ q^2)$ is the limiting value of $H(z;q)$ in the bulk liquid phase, and we have used (\ref{NEW}). Substitution of the profile (\ref{rhodecayl}) and ansatz (\ref{Hansatz}) determines immediately the coefficient $\tilde a_l(q)$ and, hence, that 
\begin{equation}
 H(z;q)=H_l(q)+\frac{\alpha\,\mu_h''(\rho_l)}{\kappa_l\,\mu_h '(\rho_l)^2}\,\frac{H_l(q)}{q^2}\;\rho'(z) +\,\cdots
 \label{Hsol}
\end{equation}
where we have simply written the decaying exponential in terms of the derivative of the density profile using $\rho'(z)\approx -\kappa_l\, \delta\rho(z)$, valid for large $z$. Using the definition (\ref{H}) to transform from $H(z;q)$ to the original $S(z;q)$, this decay is equivalent to  
\begin{equation}
 S(z;q)=S_l(q)+S_l(q)\,\frac{\mu_h''(\rho_l)}{\kappa_l\, \mu_h '(\rho_l)}\left(\!1\!+\!\frac{\alpha}{\mu_h '(\rho_l)\,q^2}\!\right)\rho'(z)\, +\,\cdots
 \label{Ssol}
\end{equation}
which is of the desired form (\ref{Slargel}), and identifies $\sigma_l(q)$ explicitly, leading to
\begin{equation}
\sigma_l(q)\;=\;\sigma_l(0)\;\frac{1+\displaystyle\frac{q^2\xi_l^2}{1+q^2R^2}}{\;1+q^2\left(R^2+\frac{R^4}{\xi_l^2}\right)\;}
\label{sigmaqsull}
\end{equation}
where again we have reinstated $R$.  An equivalent result applies for the decay of $S(z;q)$ on the gas side. The value of $\sigma_b(0)$ is also identified from (\ref{Ssol}) and can be conveniently expressed as
\begin{equation}
\sigma_b(0)\;=\, \xi_b^T\Delta\rho\,\left(\frac{\xi_b^T }{\xi_b}\right)^2\frac{\partial\rho_b}{\partial\mu}\big/\left|\frac{\partial^2\rho_b}{\partial\mu^2}\right|
\label{sigma0sull}
\end{equation}
where $\mu(\rho_b)=\mu_h(\rho_b)-\alpha\rho_b$ and $\frac{\partial\rho_b}{\partial\mu}=(\mu_h'(\rho_b)-\alpha)^{-1}$ is proportional to the bulk compressibility. It is instructive to note that only in the (artificial) limit where the range $R$  goes to zero does (\ref{sigmaqsull}) reduces to the simpler square gradient result (\ref{5}) and then the prefactor $\sigma_b(0)$ in (\ref{sigma0sull}) is given by (\ref{sigma0sgt}). More generally, (\ref{sigma0sull}) reduces to the form (\ref{sigma0sgt}) at high temperatures where the ratio $\xi_b^T/\xi_b$ is close to unity.\\

\begin{figure}[t]
\includegraphics[width=\columnwidth]{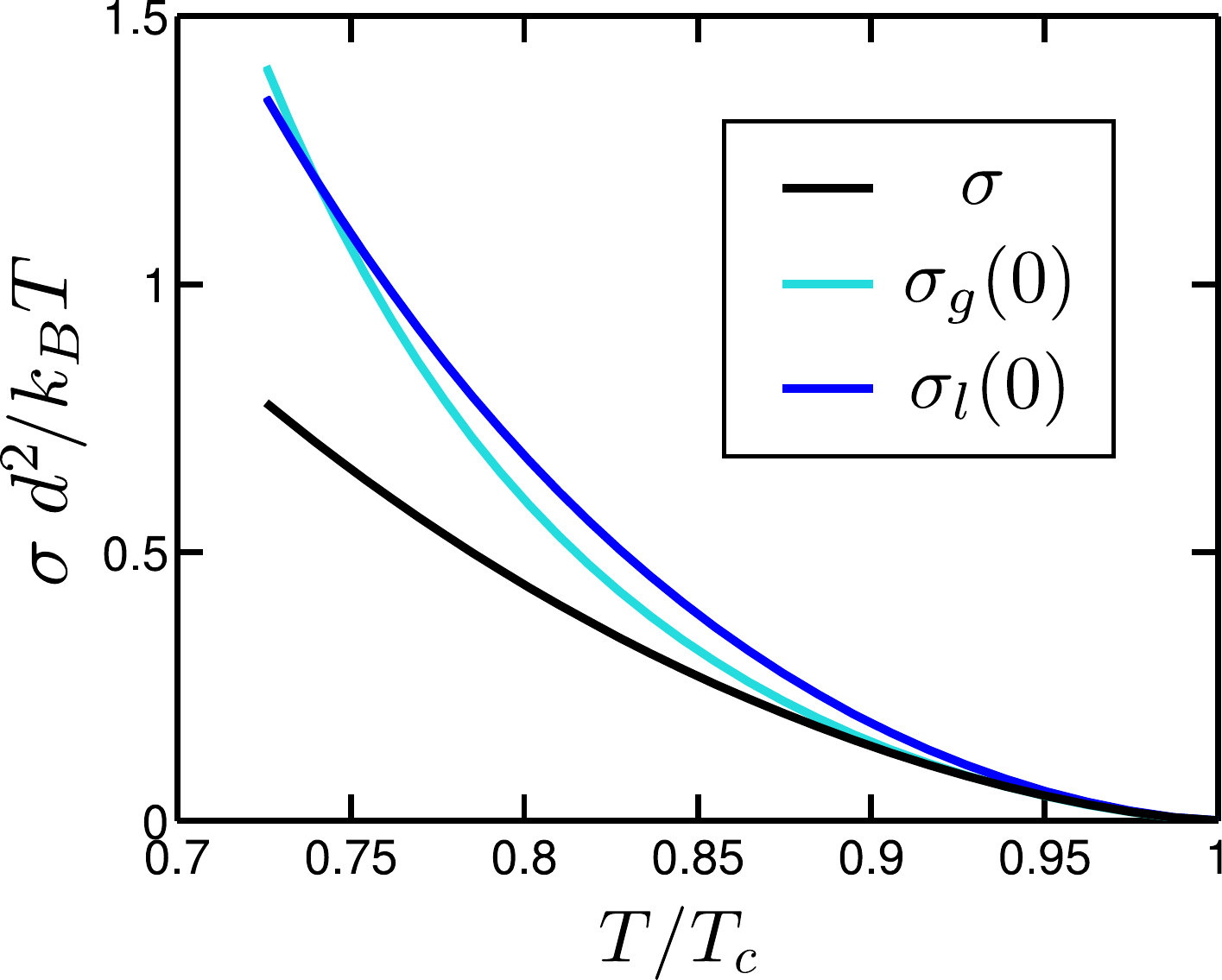}
\caption{\label{FigEx} Sullivan model results for $\sigma$, $\sigma_g(0)$ and $\sigma_l(0)$ vs. reduced temperature using the Carnahan-Starling approximation for $\mu_h(\rho)$. For $T\gtrsim 0.9\,T_c$, the three quantities are similar, in keeping with the predictions of the square-gradient theory.}
\end{figure}

Eqn.(\ref{sigmaqsull}) has a non-trivial $q$ dependence that is precisely in accord with the general prediction (\ref{LDA0}) that $q^2\sigma_b(q)\propto C_b(q)C_b(q_*)$. Again, one recognises that the numerator of (\ref{sigmaqsull}) simply echoes the wavevector dependence of the bulk direct correlation function $C_l(q)$, while the $q$ dependence in the denominator comes from that contained within $C_b(q_*)$. Plots of $\sigma_b(q)/\sigma_b(0)$ for different values of $R/\xi_b$ are shown in Fig.~\ref{Fig5}, and display the same qualitative structure as the simulation results of H\"ofling and Dietrich shown in Fig.~\ref{Fig1}. In particular, there is a pronounced maximum near $qR\approx 1$ for small values of $R/\xi_b$ corresponding to high temperatures. Expansion of $\sigma_b(q)$ determines the liquid and gas rigidity coefficients as
\begin{equation}
K_b\;=\;\sigma_b(0) \left(\,\xi_b^2-R^2-\frac{R^4}{\xi_b^2}\,\right)
\label{Ksull}
\end{equation}
At high temperatures, $\xi_b\gg R$ so that the rigidities are positive. However, at low temperatures, $\xi_b$ can be smaller than the range $R$ and (\ref{Ksull})  implies that $K_b$ can be negative. The rigidity has a change of sign when $R/\xi_b=\sqrt{(\sqrt{5}-1)/2\,}$ which, in turn, means there that is a qualitative change in the behaviour of $\sigma_b(q)$. Thus, we expect that each $K_b$ is negative at low temperatures, when the bulk correlation length is small compared to the intermolecular range $R$, but that $K_b$ is positive at high temperatures and takes values similar to those from square gradient theory. Estimates of $\xi_b/R$ at different temperatures can be taken from Lu {\it{et al}} \cite{Lu1985}. Taking $R$ to be the hard-sphere diameter, one finds that $K_l$ is negative for $T/T_c\lesssim 0.6$ and $K_g$ is negative for $T/T_c\lesssim 0.85$. We emphasise that the rigidities $K_l$ and $K_g$, identified from the asymptotic decay of $S(z;q)$. are unrelated to the effective Hamiltonian rigidity $\mathcal{K}_\textsc{\tiny ECW}^\textsc{\tiny DFT}$ described above (\ref{Kecw}). From the wavevector expansion of (\ref{Cqsull}) one sees that the fourth moment $C_4(z_1,z_2)<0$ for all finite $|z_1\!-\!z_2|$, implying that always $\mathcal{K}_\textsc{\tiny ECW}^\textsc{\tiny DFT}<0$. Thus, $\mathcal{K}_\textsc{\tiny ECW}^\textsc{\tiny DFT}$ cannot display the rich temperature dependence shown by $K_l$ and $K_g$.\\
Finally, we turn attention to the values of $\sigma_l(0)$ and $\sigma_g(0)$ given by equation (\ref{sigma0sull}) which depend on the choice of the bulk equation of state. We have determined these and also the equilibrium tension $\sigma$ given by (\ref{SullTen}), as a function of $T/T_c$, using the Carnahan-Starling approximation for the bulk hard-sphere chemical potential
 \begin{equation}
 \mu_h(\rho)\,=\,\ln\eta+\eta\;\frac{\,8-9\eta+3\eta^2}{(1-\eta)^3}
 \label{Carn}
 \end{equation}
where $\eta=\pi\rho d^3/6$ is the packing fraction, $d$ the hard-sphere diameter and recall we have set $\beta=1$. The results are shown in Fig.~\ref{FigEx}. We find that $\sigma_l(0)\approx\sigma_g(0)$ for all temperatures, and that these quantities are close to the equilibrium tension $\sigma$ for $T/T_c\gtrsim 0.9$. This is in keeping with our earlier results based on square-gradient theory with the Landau-like potential (\ref{Landau}) for which $\sigma_b(0)=\sigma$ at all temperatures. However, at lower temperatures, the values of $\sigma_b(0)$ obtained using the Carnahan-Starling approximation (\ref{Carn}) are larger than the equilibrium tension. This implies that the next-to-leading order decay of $S(z;q)$, as shown in (\ref{Scrossover}), must be included in order that $S(z;q)$ has the correct cross-over to the Wertheim Goldstone mode divergence as $q\to 0$.\\

The fact that, at low temperatures, the coefficient $\sigma_b(0)$ is substantially greater than the surface tension $\sigma$ has important implications for fluctuation effects. From (\ref{Scrossover}), we see that, for distances $|z|\gtrsim \xi_b$ and \textit{small} wavevectors in the range $\kappa_b> q> 1/\sqrt{|z|\xi_b}$, the local structure factor is well approximated by a Goldstone-mode $q^{-2}$ divergence but is considerably smaller than the Wertheim/Capillary-Wave expectation:
\begin{equation}
      S(z;q)\;\approx\; \frac{\rho'(z)\,\Delta\rho}{\beta\,q^2\sigma_b(0)} \; < \;  \frac{\rho'(z)\,\Delta\rho}{\beta\,q^2\sigma}
\end{equation}
where we have reinstated $\beta=(k_B T)^{-1}$. This reduction is not associated with any rigidity, and cannot be explained by effective Hamiltonian theory.

\begin{figure}[t]
\includegraphics[width=\columnwidth]{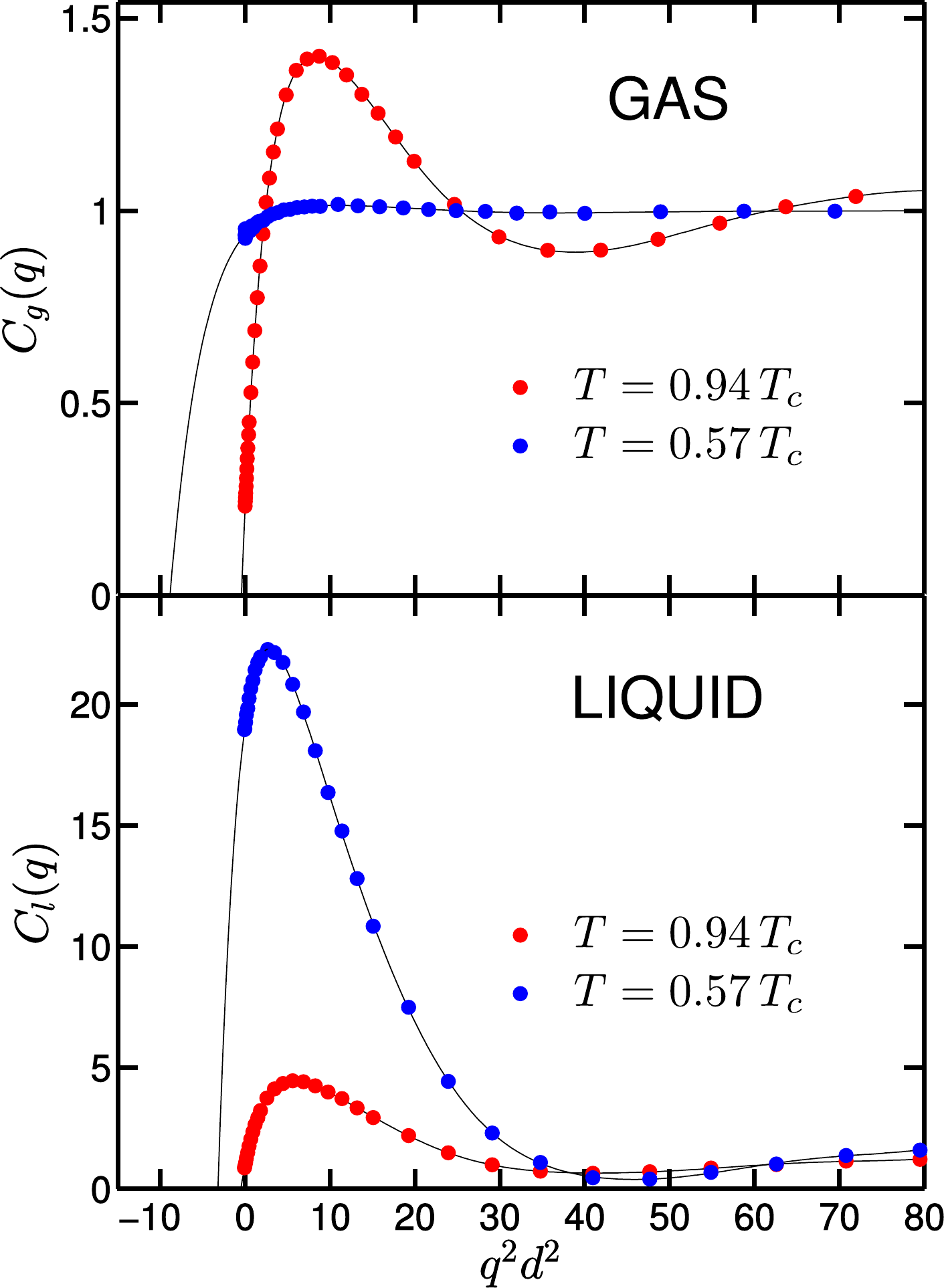}
\caption{\label{Fig6} H\"ofling and Dietrich Molecular Dynamics results for the bulk liquid and gas direct correlation functions $C_b(q)$ \cite{Hpc}, at the highest and lowest temperatures simulated. The continuous lines are polynomial fits in $q^2$ containing $15$ coefficients. The extrapolation to negative values of $q^2$ is equivalent to the analytic continuation which determines $C_b(q_*)$. See text for details. The functions $C_b(q)$ plotted here are the inverse of the conventional structure factor. Thus, these are dimensionless and correspond to $\rho_b/S_b(q)$, where $\rho_b$ is the bulk density, but this plays no role in the determination of the functions $\sigma_b(q)/\sigma_b(0)$.}
\end{figure}

\section{Comparison of theory and simulation}

\begin{figure}[t]
\includegraphics[width=\columnwidth]{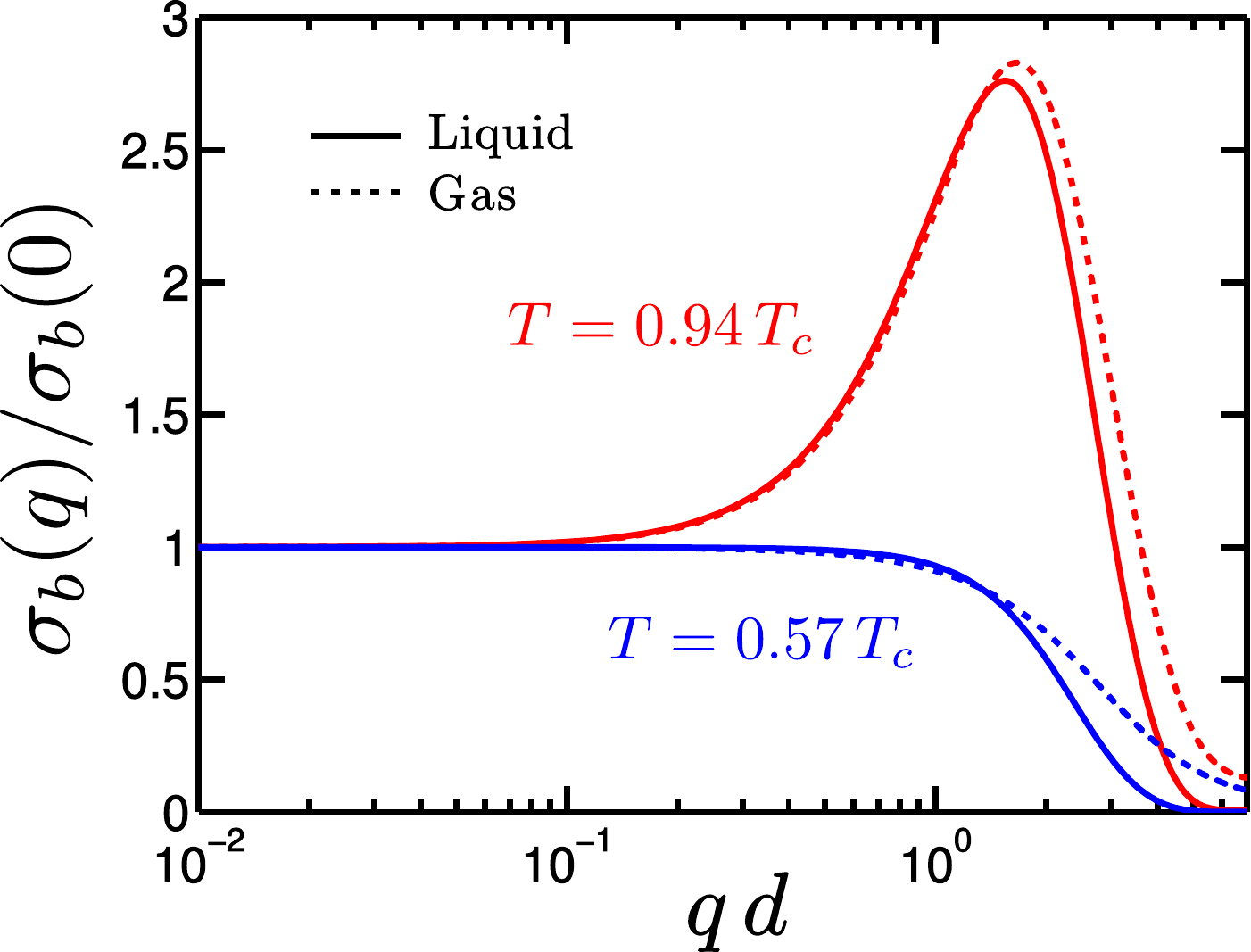}
\caption{\label{Fig7} Predicted wavevector dependent tensions $\sigma_b(q)/\sigma_b(0)$, based on the two-body approximation $q^2\sigma_b(q)\propto C_b(q)C_b(q_*)$, for the cut-off Lennard-Jones potential at $T=0.94T_c$ and $T=0.57T_c$. These employ the polynomial fits of the HD simulation results for $C_b(q)$ shown in Fig.\ref{Fig6}.}
\end{figure}

We turn now to the simulation results of H\"ofling and Dietrich (HD) \cite{Hofling2015}. These authors conducted an extensive Molecular Dynamics study of the liquid-vapour interface in a system with a Lennard-Jones potential $w(r)=-4\epsilon ((r/d)^{-6}-(r/d)^{-12})$ which is cut-off at $r_c=3.5 d$. To date, these are probably the largest simulation studies of fluctuation effects at a liquid-vapour interface involving up to half a million particles using boxes of size $L_\parallel^2 \times L_z$ with $L_\parallel=100d$ and $L_z=125d$ or $200d$. Over temperatures ranging from $T/T_c\approx 0.57$, which lies close to the triple point, up to $T/T_c\approx 0.94$, HD determine the 
bulk structure factors $S_b(q)$, the equilibrium profile $\rho(z)$, the density-density correlation function $G(z,z';q)$ and the total structure factor $S(q)$ obtained by integrating $G(z,z';q)$ over $z'$ - see equation (\ref{Stot0}).
Then, HD write $S(q)=S^{ex}(q)+S^{bg}(q)$, where $S^{bg}(q)$ is a background/bulk contribution which is {\it{approximately}} $L_gS_g(q)+L_l S_l(q)$, where $L_gL_\parallel^2$ and $L_lL_\parallel^2$ are the volumes occupied by the liquid and gas phases. More will be said of this later. From the excess contribution, HD identify a wavevector dependent tension via $\beta S^{ex}(q)=(\Delta\rho)^2/\sigma_\textsc{\tiny HD}(q)q^2$. In sharp contrast to the expectations based on extended Capillary-Wave theory, $\sigma_\textsc{\tiny HD}(q)$ has a pronounced maximum at high temperatures, although this reduces as $T$ is lowered towards the triple point (see Fig.~\ref{Fig1}). To explain this behaviour, let us focus on the properties of $\sigma_l(q)$ and $\sigma_g(q)$ using (\ref{LDA0}), which involves $C_b(q)$ and $C_b(q_*)$. As mentioned earlier, this approximation ignores any three-body contribution which is almost certainly justifiable except for large wavevectors, say $qd>\pi$. HD measured the bulk structure factors $S_b(q)$, and these immediately give us $C_b(q)=1/S_b(q)$. To determine $C_b(q_*)$, we plot $C_b(q)$ as a function of $q^2$ and, after fitting to a polynomial, extrapolate to negative $q^2$, which is equivalent to the analytic continuation. The zero of the extrapolated $C_b(q)$ occurs at $q^2=-\kappa_b^2$ and, by shifting the axis $q^2\to q^2-\kappa_b^2\equiv -q_*^2$, one obtains $C_b(q_*)$ (see Fig.~\ref{Fig6}). Note that the minimum in $C_l(q)$ at $T=0.57\, T_c$, which occurs near $qd=6.7$, corresponds to the first peak of the liquid structure factor. The results for $\sigma_l(q)/\sigma_l(0)$ and $\sigma_g(q)/\sigma_g(0)$ at $T=0.94\,T_c$ and  $T=0.57\,T_c$ are shown in Fig.~\ref{Fig7}. Qualitatively, these results are very similar to the predictions based on the Sullivan model shown in Fig.~\ref{Fig5}. In particular, at the higher temperature $T=0.94\,T_c$, we find that $\sigma_l(q)/\sigma_l(0)$ and $\sigma_g(q)/\sigma_g(0)$ are close, with each displaying a pronounced maximum at $qd\approx 1.6$. The wavevector dependence at this temperature is largely determined by the properties of the bulk structure factor so that $\sigma_b(q)/\sigma_b(0)\approx S_b(0)/S_b(q)$. At the lower temperature, however, the additional lengthscales emerging from $C_b(q_*)$ are more important and, as with the simple Sullivan model, the $\sigma_b(q)$ no longer display maxima, implying that the rigidity coefficients $K_l$ and $K_g$ are negative.\\

\begin{figure}[b]
\includegraphics[width=\columnwidth]{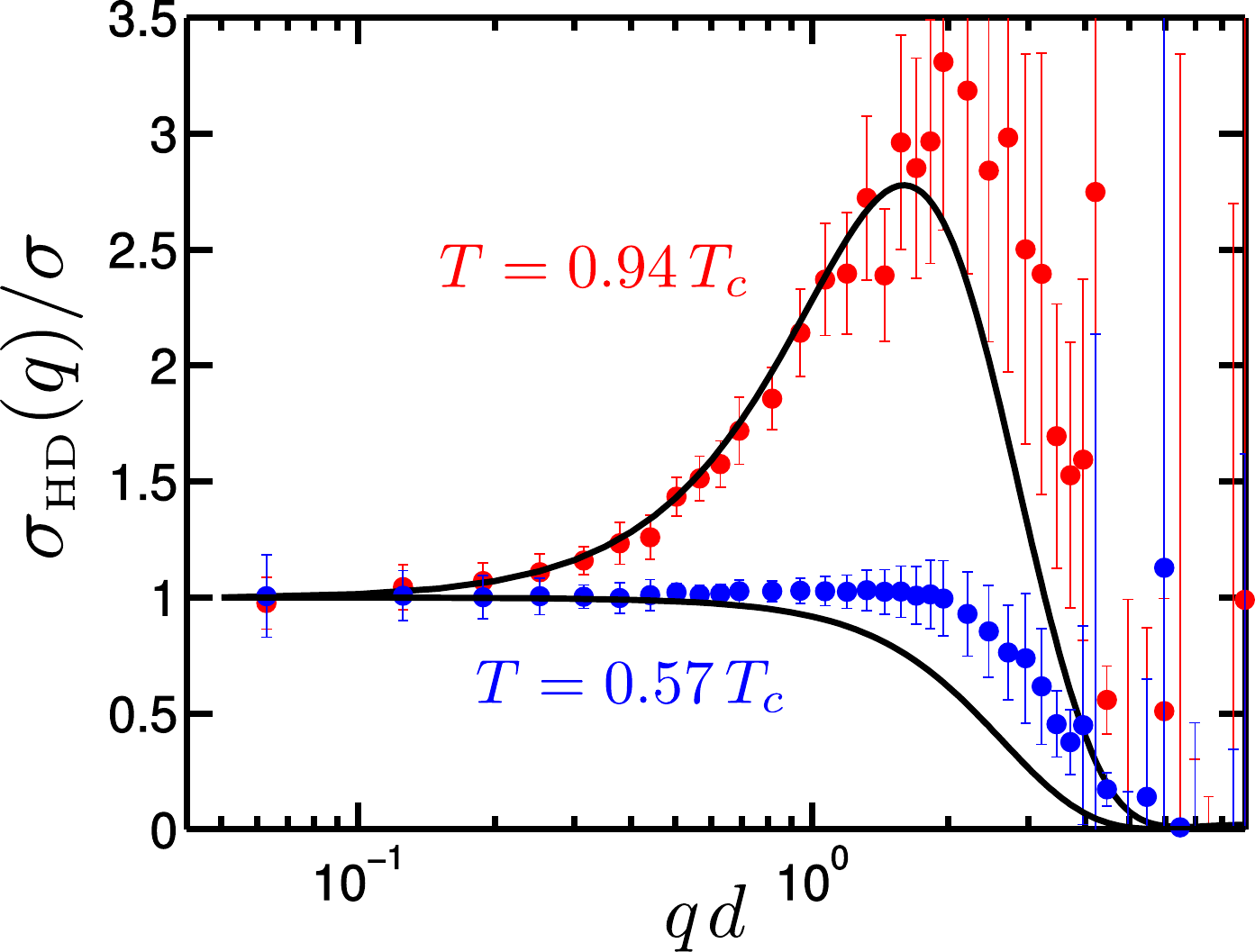}
\caption{\label{Fig8} Comparison of the H\"ofling-Dietrich simulation results \cite{Hofling2015} for $\sigma_\textup{\tiny HD}(q)$ (circles) with the prediction for $\sigma_\textup{wt}(q)$ based on (\ref{comp}) (continuous lines), for $T/T_c=0.94$ and $0.57$. The fractions $f_g=0.49$ ($T=0.94\,T_c$) and $f_g=0.47$ ($T=0.57\,T_c$) are obtained from the Molecular Dynamics density profiles \cite{Hpc}, although using $f_g=0.5$ produces similar results.}
\end{figure}

To make more quantitative comparison with the simulation results, we use the result (\ref{StotQM}) for the total structure factor that we obtained for the double-quartic potential. This result suggests that if the contribution from the (unknown) microscopic lengthscale $\zeta(q)$ is negligible, as it is in the double-quartic model, then the weighted combination
\begin{equation}
\frac{\sigma_\textsc{\tiny wt}(0)}{\sigma_\textsc{\tiny wt}(q)}\;=\; f_g\,\frac{\sigma_{g}(0)}{\sigma_{g}(q)}\;+\;f_l\,\frac{\sigma_{l}(0)}{\sigma_{l}(q)}
\label{comp}
\end{equation}
should be a good approximation to $\sigma_\textup{\tiny HD}(0)/\sigma_\textup{\tiny HD}(q)$. Here, the fractions $f_g$ and $f_l$ are determined via (\ref{f}) from the simulation density profiles \cite{Hpc}, and are very close to $1/2$ at both temperatures. In the simulations, it was found that at high temperatures $\sigma_\textsc{\tiny HD}(q)$ approaches the, independently determined, equilibrium tension $\sigma$, as $q\to 0$. However, at low temperatures, there is considerable error in determining $\sigma_\textsc{\tiny HD}(0)$. In Fig.~\ref{Fig1} (Fig.~3(b) of \cite{Hofling2015}), it is assumed that $\sigma_\textsc{\tiny HD}(0)=\sigma$. One should be aware that the simulation results for $\sigma_\textsc{\tiny HD}(q)$ are sensitive, particularly at lower temperatures, to what is deemed to be the background contribution to the total structure factor $S(q)$. HD settled upon a $S^{bg}(q)$ obtained by integrating a "background" pair correlation function $G^{bg}(z,z';q)$ approximated by its bulk values except when $zz'<0$, in which case it is set to zero. The comparison of the approximation (\ref{comp}) with the HD simulation results for $ \sigma_\textsc{\tiny HD}(q)$ is shown in Fig.~\ref{Fig8}, using $\sigma_l(q)$ and $\sigma_g(q)$ given in Fig.~\ref{Fig7}. At $T=0.94\,T_c$, there is remarkable agreement between the theoretical prediction (\ref{comp}) and the simulation results. 
At $T=0.57\,T_c$, the theoretical expression (\ref{comp}) captures correctly the observed flattening of $\sigma_\textsc{\tiny HD}(q)$. This occurs because the rigidities characterising $\sigma_l(q)$ and $\sigma_g(q)$ are both negative at this lowest temperature. In this case, the agreement is not quite as quantitative as at the higher temperature. There are a number of reasons for this: a) At lower temperatures, the results for $\sigma_\textsc{\tiny HD}(q)$ are more sensitive to the choice of background $S^{bg}(q)$ used by HD \cite{Hpc}. b) As we observed for the Sullivan model, the coefficients $\sigma_b(0)$ may deviate from $\sigma$ at lower temperatures. This would mean that one would also have to include in (\ref{comp}) contributions from the next-to-leading order terms in $S(z;q)$, shown in (\ref{Scrossover}). c) Three-body effects contained in $C_b^{(3)}(q;i\kappa_b,0)$, omitted in our treatment, may be more important. d) Our predictions may be affected by our going well below the Fisher-Widom temperature $T_{FW}$, although inspection of the simulation density profiles gives no indication of any oscillatory behaviour \cite{Hpc}. Nevertheless, even at the lower temperature $T=0.57\,T_c$, the agreement is certainly at least semi-quantitative and explains the absence of the pronounced maximum in $\sigma_\textsc{\tiny HD}(q)$ observed at the highest temperature.

\section{Conclusions}

In summary, we have shown that, for systems with short-ranged intermolecular forces, the local structure factor splits unambiguously into bulk and interfacial contributions {\it{far}} from a liquid-gas interface. One can characterise the latter by distinct liquid and gas wavevector dependent tensions, $\sigma_l(q)$ and $\sigma_g(q)$, which are themselves determined solely by bulk two-point, and three-point correlation functions. The tensions $\sigma_l(q)$ and $\sigma_g(q)$, and any weighted combination of these, are not to be interpreted as describing the energy cost of interfacial fluctuations. However, using these functions we can readily explain the findings of recent Molecular Dynamics simulations of the total structure factor for a free interface in a system with cut-off Lennard-Jones forces (See Figs.~\ref{Fig1}, \ref{Fig5}, \ref{Fig7}, and \ref{Fig8}). The rigidities $K_l$ and $K_g$ defined from $\sigma_l(q)$ and $\sigma_g(q)$ are positive at high $T$ but become negative at lower temperatures (approaching the bulk triple point), when the (Ornstein-Zernike) bulk correlation length $\xi_b$ is small in comparison with the range of the intermolecular potential (or molecular diameter).\\

Our analysis shows that one may determine the asymptotic decay of $S(z;q)$ as $|z|\to\infty$ in terms of the bulk structure factors $S_b(q)$, the equilibrium density profile $\rho(z)$ and the new wavevector dependent tensions $\sigma_b(q)$, which depend on only two- and three-point bulk direct correlation functions. Our theory does not provide any new prescriptive means of determining the broadening of the interfacial profile $\rho(z)$ arising from interfacial fluctuations. The traditional, 
semi-heuristic way of predicting this is to use the standard Capillary-Wave Hamiltonian, with an ad-hoc cutoff $q_{max}$, and to
convolute an "intrinsic profile", similar to that predicted by mean-field theory, with a probability distribution function for the interfacial position. One might arguably use an extended Capillary-Wave model,
with a wavevector dependent surface tension $\sigma_\textup{\tiny ECW}(q)$,
for this purpose, although such an approach is far from rigorous and is known to omit non-local effects \cite{Fernandez2013}. However, even if one insisted on using an extended Capillary-Wave theory to predict the equilibrium density profile $\rho(z)$, the {\it{wavevector dependence}} extracted from $S(z;q)$  must be characterised by the $\sigma_b(q)$. We emphasise that this can be predicted without resort to effective Hamiltonian theory and is \textit{unrelated} to any $\sigma_\textup{\tiny ECW}(q)$. It follows that extended Capillary-Wave models fail to describe the properties of the microscopic local structure factor, implying that they are not a useful means to model the fluctuations of the interfacial region when $q$ is comparable with the inverse bulk correlation lengths.\\

It would be interesting to extend the present study to long-ranged dispersion forces, for which the intermolecular potential decays as $w(r)\approx-\alpha/r^6$. Extended Capillary-Wave models derived from DFT predict that $\sigma_\textsc{\tiny ECW}^\textsc{\tiny DFT}=\sigma+A\, q^2\ln(qb)+\cdots$, where $A\propto\alpha(\Delta \rho)^2$ and $b$ is a microscopic cut-off associated with the potential \cite{Napiorkowski1993}. We suspect that such a non-analytic term appears in the corrections to the Goldstone-mode divergence (\ref{SGM}) of the local structure factor $S(z;q)$ or total structure factor $S(q)$ \cite{Parry2014}. Beyond this, however, we anticipate that approaches based on extended Capillary-Wave models fail for the same reasons described here: namely, they do not distinguish between the structure of $S(z;q)$ either side of the interface, and predict the wrong sign for rigidity-like corrections. These should be positive at high temperatures, very similar to systems with short-ranged forces.\\

For the majority of this paper, we have deliberately avoided the close proximity of the bulk critical point, and have mentioned only that, in three dimensions ($d=3$), we expect that the rigidity coefficients $K_b$ approach the same positive value as $T\to T_c^-$. In concluding our article, we estimate the critical value of $K_b$ by adapting the square-gradient theory of Sec.~III. To allow for non-classical critical singularities, we follow the approach of Fisk and Widom \cite{Rowlinson1982,Fisk1969} which replaces the mean-field Landau free-energy density (\ref{Landau}) with 
\begin{equation}\label{Fisk}
 \phi(\rho)=-\frac{\,t^\gamma}{2\,}\,(\rho-\rho_c)^2\,+\,\frac{u}{\delta+1}\,(\rho-\rho_c)^{\,\delta+1}
\end{equation}
and also sets $f_2\propto t^{-\eta\nu}$. Here $\gamma$, $\delta$, $\eta$, and $\nu$ (and also $\alpha$, below) are standard bulk critical exponents, the values of which are input for this phenomenological treatment of bulk critical effects. In place of the mean-field result $\sigma\propto t^{3/2}$, the Fisk-Widom theory predicts $\sigma\propto t^{\tilde\mu}$, with the critical exponent $\tilde\mu$ determined correctly as $\tilde\mu=2-\alpha-\nu$ which simplifies to $\tilde\mu=(d-1)\nu$ for $d<4$ on using hyperscaling. The Fisk-Widom theory is also believed to predict accurately the value of the universal critical amplitude $B_\textsc{\tiny FW}$ appearing in $\sigma=B_\textsc{\tiny FW}f_2(\Delta\rho)^2/\xi_b$ (referred to as $K$ in section 9.4 of \cite{Rowlinson1982}). As stated earlier, within mean-field approximation (or equivalently for $d>4$), this amplitude is $B_\textsc{\tiny FW}=1/6$, while in dimension $d=3$, the Fisk-Widom theory predicts that it takes the non-classical value $B_\textsc{\tiny FW}\approx 1/7$. This prediction is in good agreement with independent estimates based on RG analysis employing the $\epsilon$ expansion \cite{Ohta1977}. Using the Fisk-Widom free-energy density (\ref{Fisk}) and (\ref{3}) (or equivalently (\ref{sigma0sgt})), we find that $\sigma_b(0)\propto t ^{\tilde\mu}$, implying that the coefficient $\sigma_b(0)$ displays the {\it{same}} critical singularity as the equilibrium surface tension $\sigma$ in all dimensions as $T\to T_c^-$.
The ratio $\sigma_b(0)/\sigma$ is easily obtained
\begin{equation}
\frac{\sigma_b(0)}{\sigma}\;=\;\frac{1}{\,2\,\delta\,B_\textsc{\tiny FW}}
\end{equation}
Taking $\delta=5$, consistent with the Fisk-Widom parametrization, this ratio is approximately $0.7$ whereas in the mean-field square gradient treatment using the Landau form (\ref{Landau}), we have $\delta=3$, $B_\textsc{\tiny FW}=1/6$, and the ratio is unity as stated previously --see (\ref{6}). Then, using the result $K_b=\sigma_b(0)\,\xi_b^2$, which follows directly from (\ref{5}), we can express the critical value of the rigidity coefficient as 
\begin{equation}
\frac{K_b}{k_B T_c}\;=\;\frac{1}{8\pi\,\delta\, B_\textsc{\tiny FW}\,\omega_c}
\label{Fisk2}
\end{equation}
where $\omega_c$ is the universal critical value of the dimensionless 'wetting' parameter $\omega$, defined after (\ref{Scaling}). Using the same values of $\delta$ and $B_\textsc{\tiny FW}$, and $\omega_c\approx 0.87$ \cite{Das2011}, we obtain $K_b/k_B T_c\approx (15)^{-1}$ which may be tested in simulation studies similar to those of refs. \cite{Hofling2015,Das2011}.\\

\acknowledgments

We are extremely grateful to Felix H\"ofling for providing his
simulation results to us, and for helpful
discussions.  Correspondence with Pedro Tarazona, Enrique Chac\'{o}n and Eva Fern\'{a}ndez provided further stimulation. AOP acknowledges the EPSRC, UK for grant EP/J009636/1. CR acknowledges partial support from "EPSRC Mathematics Platform" under grant EP/I019111/1.

\section{Appendix; The structure factor within Capillary-Wave theory}

The expression for the local structure factor within Capillary-wave theory follows directly from the analysis of Bedeaux and Weeks \cite{Bedeaux1985}. These authors elegantly determined the equilibrium density profile $\rho(z)$ and the pair correlation function
$G(z,z';q)$ for an interface in a gravitational field in any dimension $d$, under the assumption that the density operator is a simple step function
\begin{equation}
\hat\rho({\bf{r}})=\rho_g+\Delta\rho\,\theta(z-\ell({\bf{x}}))
\end{equation}
For the profile, they obtain the error function
\begin{equation}
\rho_\textup{\tiny CW}(z)=\rho_l-\frac{2\Delta\rho}{\sqrt\pi}\int_\tau^\infty e^{-x^2} dx
\end{equation}
where $\tau=z/2\xi_\perp$ is a scaling variable. Of course such scaling is strictly applicable in dimension $d<3$, where fluctuations dominate and one can consider the limit $z\to \infty$ and that the interfacial width $\xi_\perp\to \infty$ with $\tau$ left arbitrary. In this regime, Bedeaux and Weeks also determine the scaling of the pair correlation function and its Fourier transform $G_\textup{\tiny CW}(z,z';q)=\xi_\parallel^2\mathcal{G}(\tau,\tau';Q)$ as
\begin{equation}
\frac{\mathcal{G}(\tau,\tau';Q)}{2\sqrt\pi(\Delta\rho)^2}=e^{-\tau^2/2}e^{-{\tau'}^2/2}\sum_{m=0}^\infty\frac{1}{\hat f_{m+1}(Q)} \psi_m(\tau)\psi_m(\tau')
\label{Gscale}
\end{equation}
where $Q=q\,\xi_\parallel$ and $\hat f_m(Q)$ is the Fourier transform of the $m$-th power of the height-height correlation function $\langle\ell({\bf{x}})\ell(0)\rangle/\xi_\perp^2$. The eigen-functions in this expansions are those of the harmonic oscillator
\begin{equation}
\psi_n(\tau)=H_n(\tau)e^{-\tau^2/2}(\sqrt\pi 2^n n!)^{-1/2}
\label{Hn}
\end{equation}
where $H_n(\tau)$ are simple Hermite polynomials
\begin{equation}
(-1)^n\frac{d^n}{d\tau^n}e^{-\tau^2}=H_n(\tau)e^{-\tau^2}
\label{orthog}
\end{equation}
To determine the capillary-wave expression for $S(z;q)$ we simply integrate (\ref{Gscale}) over $\tau'$. The orthogonality condition cancels all but the ground state contribution leaving
\begin{equation}
\int d\tau'\mathcal{G}(\tau,\tau';Q)=\frac{2(\Delta\rho^2)}{\sqrt\pi}\frac{e^{-\tau^2/2}}{1+Q^2}
\end{equation}
where we have substituted $\hat f_1(Q)=1/(1+Q^2)$, equivalent to the expression (\ref{CW}) for the height-height correlation function. In the original variables, this expression becomes
\begin{equation}
\beta S_\textup{\tiny CW}(z;q)= \frac{\rho'(z)\Delta \rho}{mg\Delta\rho+\sigma q^2}
\end{equation}
The RHS is precisely  Wertheim's single eigenfunction result (\ref{SGM}). That Capillary-Wave theory yields solely the Goldstone mode $q^{-2}$ term, when $g=0$, is perhaps not surprising. But it is pleasing that this result emerges from an explicit analysis. \\

\bibliography{wetting}

\begin{thebibliography}{44}
\expandafter\ifx\csname natexlab\endcsname\relax\def\natexlab#1{#1}\fi
\expandafter\ifx\csname bibnamefont\endcsname\relax
  \def\bibnamefont#1{#1}\fi
\expandafter\ifx\csname bibfnamefont\endcsname\relax
  \def\bibfnamefont#1{#1}\fi
\expandafter\ifx\csname citenamefont\endcsname\relax
  \def\citenamefont#1{#1}\fi
\expandafter\ifx\csname url\endcsname\relax
  \def\url#1{\texttt{#1}}\fi
\expandafter\ifx\csname urlprefix\endcsname\relax\def\urlprefix{URL }\fi
\providecommand{\bibinfo}[2]{#2}
\providecommand{\eprint}[2][]{\url{#2}}

\bibitem[{\citenamefont{Buff et~al.}(1965)\citenamefont{Buff, Lovett, and
  Stillinger}}]{BLS1965}
\bibinfo{author}{\bibfnamefont{F.~P.} \bibnamefont{Buff}},
  \bibinfo{author}{\bibfnamefont{R.~A.} \bibnamefont{Lovett}},
  \bibnamefont{and} \bibinfo{author}{\bibfnamefont{F.~H.}
  \bibnamefont{Stillinger}}, \bibinfo{journal}{Phys. Rev. Lett.}
  \textbf{\bibinfo{volume}{15}}, \bibinfo{pages}{621} (\bibinfo{year}{1965}).

\bibitem[{\citenamefont{Weeks}(1977)}]{Weeks1977}
\bibinfo{author}{\bibfnamefont{J.~D.} \bibnamefont{Weeks}},
  \bibinfo{journal}{J. Chem. Phys.} \textbf{\bibinfo{volume}{67}},
  \bibinfo{pages}{3106} (\bibinfo{year}{1977}).

\bibitem[{\citenamefont{Bedeaux and Weeks}(1985)}]{Bedeaux1985}
\bibinfo{author}{\bibfnamefont{D.}~\bibnamefont{Bedeaux}} \bibnamefont{and}
  \bibinfo{author}{\bibfnamefont{J.~D.} \bibnamefont{Weeks}},
  \bibinfo{journal}{J. Chem. Phys.} \textbf{\bibinfo{volume}{82}},
  \bibinfo{pages}{972} (\bibinfo{year}{1985}).

\bibitem[{\citenamefont{Wertheim}(1976)}]{Wertheim1976}
\bibinfo{author}{\bibfnamefont{M.~S.} \bibnamefont{Wertheim}},
  \bibinfo{journal}{J. Chem. Phys.} \textbf{\bibinfo{volume}{65}},
  \bibinfo{pages}{2377} (\bibinfo{year}{1976}).

\bibitem[{\citenamefont{Triezenberg and Zwanzig}(1972)}]{Triezenberg1972}
\bibinfo{author}{\bibfnamefont{D.~G.} \bibnamefont{Triezenberg}}
  \bibnamefont{and} \bibinfo{author}{\bibfnamefont{R.}~\bibnamefont{Zwanzig}},
  \bibinfo{journal}{Phys. Rev. Lett.} \textbf{\bibinfo{volume}{28}},
  \bibinfo{pages}{1183} (\bibinfo{year}{1972}).

\bibitem[{\citenamefont{Aarts et~al.}(2004)\citenamefont{Aarts, Schmidt, and
  Lekkerkerker}}]{Aarts2004}
\bibinfo{author}{\bibfnamefont{D.~G. A.~L.} \bibnamefont{Aarts}},
  \bibinfo{author}{\bibfnamefont{M.}~\bibnamefont{Schmidt}}, \bibnamefont{and}
  \bibinfo{author}{\bibfnamefont{H.~N.~W.} \bibnamefont{Lekkerkerker}},
  \bibinfo{journal}{Science} \textbf{\bibinfo{volume}{304}},
  \bibinfo{pages}{847} (\bibinfo{year}{2004}).

\bibitem[{\citenamefont{H{\"o}fling and Dietrich}(2015)}]{Hofling2015}
\bibinfo{author}{\bibfnamefont{F.}~\bibnamefont{H{\"o}fling}} \bibnamefont{and}
  \bibinfo{author}{\bibfnamefont{S.}~\bibnamefont{Dietrich}},
  \bibinfo{journal}{Europhys. Lett.} \textbf{\bibinfo{volume}{109}},
  \bibinfo{pages}{46002} (\bibinfo{year}{2015}).

\bibitem[{\citenamefont{Zia}(1985)}]{Zia1985}
\bibinfo{author}{\bibfnamefont{R.~K.~P.} \bibnamefont{Zia}},
  \bibinfo{journal}{Nuclear Phys. B} \textbf{\bibinfo{volume}{251}},
  \bibinfo{pages}{676} (\bibinfo{year}{1985}).

\bibitem[{\citenamefont{Romero-Roch{\'i}n
  et~al.}(1991)\citenamefont{Romero-Roch{\'i}n, Varea, and
  Robledo}}]{Rochin1991}
\bibinfo{author}{\bibfnamefont{V.}~\bibnamefont{Romero-Roch{\'i}n}},
  \bibinfo{author}{\bibfnamefont{C.}~\bibnamefont{Varea}}, \bibnamefont{and}
  \bibinfo{author}{\bibfnamefont{A.}~\bibnamefont{Robledo}},
  \bibinfo{journal}{Phys. Rev. A} \textbf{\bibinfo{volume}{44}},
  \bibinfo{pages}{8417} (\bibinfo{year}{1991}).

\bibitem[{\citenamefont{Blokhuis and Bedeaux}(1993)}]{Blokhuis1993}
\bibinfo{author}{\bibfnamefont{E.~M.} \bibnamefont{Blokhuis}} \bibnamefont{and}
  \bibinfo{author}{\bibfnamefont{D.}~\bibnamefont{Bedeaux}},
  \bibinfo{journal}{Mol. Phys.} \textbf{\bibinfo{volume}{80}},
  \bibinfo{pages}{705} (\bibinfo{year}{1993}).

\bibitem[{\citenamefont{Napi{\'o}rkowski and
  Dietrich}(1993)}]{Napiorkowski1993}
\bibinfo{author}{\bibfnamefont{M.}~\bibnamefont{Napi{\'o}rkowski}}
  \bibnamefont{and} \bibinfo{author}{\bibfnamefont{S.}~\bibnamefont{Dietrich}},
  \bibinfo{journal}{Phys. Rev. E} \textbf{\bibinfo{volume}{47}},
  \bibinfo{pages}{1836} (\bibinfo{year}{1993}).

\bibitem[{\citenamefont{Parry and Boulter}(1994)}]{Parry1994}
\bibinfo{author}{\bibfnamefont{A.~O.} \bibnamefont{Parry}} \bibnamefont{and}
  \bibinfo{author}{\bibfnamefont{C.~J.} \bibnamefont{Boulter}},
  \bibinfo{journal}{J. Phys.: Condens. Matter} \textbf{\bibinfo{volume}{6}},
  \bibinfo{pages}{7199} (\bibinfo{year}{1994}).

\bibitem[{\citenamefont{Mecke and Dietrich}(1999)}]{Mecke1999}
\bibinfo{author}{\bibfnamefont{K.~R.} \bibnamefont{Mecke}} \bibnamefont{and}
  \bibinfo{author}{\bibfnamefont{S.}~\bibnamefont{Dietrich}},
  \bibinfo{journal}{Phys. Rev. E} \textbf{\bibinfo{volume}{59}},
  \bibinfo{pages}{6766} (\bibinfo{year}{1999}).

\bibitem[{\citenamefont{Blokhuis et~al.}(1999)\citenamefont{Blokhuis,
  Groenewold, and Bedeaux}}]{Blokhuis1999}
\bibinfo{author}{\bibfnamefont{E.~M.} \bibnamefont{Blokhuis}},
  \bibinfo{author}{\bibfnamefont{J.}~\bibnamefont{Groenewold}},
  \bibnamefont{and} \bibinfo{author}{\bibfnamefont{D.}~\bibnamefont{Bedeaux}},
  \bibinfo{journal}{Mol. Phys.} \textbf{\bibinfo{volume}{96}},
  \bibinfo{pages}{397} (\bibinfo{year}{1999}).

\bibitem[{\citenamefont{Fradin et~al.}(2000)\citenamefont{Fradin, Braslau,
  Luzet, Smilgies, Alba, Boudet, Mecke, and Daillant}}]{Fradin2000}
\bibinfo{author}{\bibfnamefont{C.}~\bibnamefont{Fradin}},
  \bibinfo{author}{\bibfnamefont{A.}~\bibnamefont{Braslau}},
  \bibinfo{author}{\bibfnamefont{D.}~\bibnamefont{Luzet}},
  \bibinfo{author}{\bibfnamefont{D.}~\bibnamefont{Smilgies}},
  \bibinfo{author}{\bibfnamefont{M.}~\bibnamefont{Alba}},
  \bibinfo{author}{\bibfnamefont{N.}~\bibnamefont{Boudet}},
  \bibinfo{author}{\bibfnamefont{K.}~\bibnamefont{Mecke}}, \bibnamefont{and}
  \bibinfo{author}{\bibfnamefont{J.}~\bibnamefont{Daillant}},
  \bibinfo{journal}{Nature} \textbf{\bibinfo{volume}{403}},
  \bibinfo{pages}{871} (\bibinfo{year}{2000}).

\bibitem[{\citenamefont{Mora et~al.}(2003)\citenamefont{Mora, Daillant, Mecke,
  Luzet, Braslau, Alba, and Struth}}]{Mora2003}
\bibinfo{author}{\bibfnamefont{S.}~\bibnamefont{Mora}},
  \bibinfo{author}{\bibfnamefont{J.}~\bibnamefont{Daillant}},
  \bibinfo{author}{\bibfnamefont{K.}~\bibnamefont{Mecke}},
  \bibinfo{author}{\bibfnamefont{D.}~\bibnamefont{Luzet}},
  \bibinfo{author}{\bibfnamefont{A.}~\bibnamefont{Braslau}},
  \bibinfo{author}{\bibfnamefont{M.}~\bibnamefont{Alba}}, \bibnamefont{and}
  \bibinfo{author}{\bibfnamefont{B.}~\bibnamefont{Struth}},
  \bibinfo{journal}{Phys. Rev. Lett.} \textbf{\bibinfo{volume}{90}},
  \bibinfo{pages}{216101} (\bibinfo{year}{2003}).

\bibitem[{\citenamefont{Vink et~al.}(2005)\citenamefont{Vink, Horbach, and
  Binder}}]{Vink2005}
\bibinfo{author}{\bibfnamefont{R.~L.~C.} \bibnamefont{Vink}},
  \bibinfo{author}{\bibfnamefont{J.}~\bibnamefont{Horbach}}, \bibnamefont{and}
  \bibinfo{author}{\bibfnamefont{K.}~\bibnamefont{Binder}},
  \bibinfo{journal}{J. Chem. Phys.} \textbf{\bibinfo{volume}{122}},
  \bibinfo{pages}{134905} (\bibinfo{year}{2005}).

\bibitem[{\citenamefont{Blokhuis et~al.}(2008)\citenamefont{Blokhuis, Kuipers,
  and Vink}}]{Blokhuis2008}
\bibinfo{author}{\bibfnamefont{E.~M.} \bibnamefont{Blokhuis}},
  \bibinfo{author}{\bibfnamefont{J.}~\bibnamefont{Kuipers}}, \bibnamefont{and}
  \bibinfo{author}{\bibfnamefont{R.~L.~C.} \bibnamefont{Vink}},
  \bibinfo{journal}{Phys. Rev. Lett.} \textbf{\bibinfo{volume}{101}},
  \bibinfo{pages}{086101} (\bibinfo{year}{2008}).

\bibitem[{\citenamefont{Blokhuis}(2009)}]{Blokhuis2009}
\bibinfo{author}{\bibfnamefont{E.~M.} \bibnamefont{Blokhuis}},
  \bibinfo{journal}{J. Chem. Phys.} \textbf{\bibinfo{volume}{130}},
  \bibinfo{pages}{014706} (\bibinfo{year}{2009}).

\bibitem[{\citenamefont{Chac{\'o}n et~al.}(2014)\citenamefont{Chac{\'o}n,
  Fern{\'a}ndez, and Tarazona}}]{Chacon2014}
\bibinfo{author}{\bibfnamefont{E.}~\bibnamefont{Chac{\'o}n}},
  \bibinfo{author}{\bibfnamefont{E.~M.} \bibnamefont{Fern{\'a}ndez}},
  \bibnamefont{and} \bibinfo{author}{\bibfnamefont{P.}~\bibnamefont{Tarazona}},
  \bibinfo{journal}{Phys. Rev. E} \textbf{\bibinfo{volume}{89}},
  \bibinfo{pages}{042406} (\bibinfo{year}{2014}).

\bibitem[{\citenamefont{Parry et~al.}(2014)\citenamefont{Parry, Rasc{\'o}n,
  Willis, and Evans}}]{Parry2014}
\bibinfo{author}{\bibfnamefont{A.~O.} \bibnamefont{Parry}},
  \bibinfo{author}{\bibfnamefont{C.}~\bibnamefont{Rasc{\'o}n}},
  \bibinfo{author}{\bibfnamefont{G.}~\bibnamefont{Willis}}, \bibnamefont{and}
  \bibinfo{author}{\bibfnamefont{R.}~\bibnamefont{Evans}}, \bibinfo{journal}{J.
  Phys.: Condens. Matter} \textbf{\bibinfo{volume}{26}},
  \bibinfo{pages}{355008} (\bibinfo{year}{2014}).

\bibitem[{\citenamefont{Parry et~al.}(2015)\citenamefont{Parry, Rasc{\'o}n, and
  Evans}}]{Parry2015}
\bibinfo{author}{\bibfnamefont{A.~O.} \bibnamefont{Parry}},
  \bibinfo{author}{\bibfnamefont{C.}~\bibnamefont{Rasc{\'o}n}},
  \bibnamefont{and} \bibinfo{author}{\bibfnamefont{R.}~\bibnamefont{Evans}},
  \bibinfo{journal}{Phys. Rev. E} \textbf{\bibinfo{volume}{91}},
  \bibinfo{pages}{030401(R)} (\bibinfo{year}{2015}).

\bibitem[{\citenamefont{Evans}(1979)}]{Evans1979}
\bibinfo{author}{\bibfnamefont{R.}~\bibnamefont{Evans}}, \bibinfo{journal}{Adv.
  Phys.} \textbf{\bibinfo{volume}{28}}, \bibinfo{pages}{143}
  (\bibinfo{year}{1979}).

\bibitem[{\citenamefont{Rowlinson and Widom}(1982)}]{Rowlinson1982}
\bibinfo{author}{\bibfnamefont{J.~S.} \bibnamefont{Rowlinson}}
  \bibnamefont{and} \bibinfo{author}{\bibfnamefont{B.}~\bibnamefont{Widom}},
  \emph{\bibinfo{title}{{Molecular Theory of Capillarity}}}
  (\bibinfo{publisher}{Clarendon Press}, \bibinfo{year}{1982}).

\bibitem[{\citenamefont{Evans}(1990)}]{Evans1990}
\bibinfo{author}{\bibfnamefont{R.}~\bibnamefont{Evans}}, in
  \emph{\bibinfo{booktitle}{{Liquids at interfaces}}}, edited by
  \bibinfo{editor}{\bibfnamefont{J.}~\bibnamefont{Charvolin}},
  \bibinfo{editor}{\bibfnamefont{J.~F.} \bibnamefont{Joanny}},
  \bibnamefont{and}
  \bibinfo{editor}{\bibfnamefont{J.}~\bibnamefont{Zinn-Justin}}
  (\bibinfo{publisher}{Elsevier}, \bibinfo{year}{1990}), p.~\bibinfo{pages}{1}.

\bibitem[{\citenamefont{Hansen and McDonald}(2006)}]{Hansen2006}
\bibinfo{author}{\bibfnamefont{J.~P.} \bibnamefont{Hansen}} \bibnamefont{and}
  \bibinfo{author}{\bibfnamefont{I.~R.} \bibnamefont{McDonald}},
  \emph{\bibinfo{title}{{Theory of Simple Liquids}}}
  (\bibinfo{publisher}{Academic Press}, \bibinfo{year}{2006}),
  \bibinfo{edition}{3rd} ed.

\bibitem[{\citenamefont{Henderson}(1992)}]{Henderson1992}
\bibinfo{author}{\bibfnamefont{J.~R.} \bibnamefont{Henderson}}, in
  \emph{\bibinfo{booktitle}{{Fundamentals of Inhomogeneous Fluids}}}, edited by
  \bibinfo{editor}{\bibfnamefont{D.}~\bibnamefont{Henderson}}
  (\bibinfo{publisher}{Marcel {D}ekker, {I}nc.}, \bibinfo{year}{1992}),
  chap.~\bibinfo{chapter}{2}.

\bibitem[{\citenamefont{Henderson}(1987)}]{Henderson1987}
\bibinfo{author}{\bibfnamefont{J.~R.} \bibnamefont{Henderson}},
  \bibinfo{journal}{Phys. Rev. A} \textbf{\bibinfo{volume}{36}},
  \bibinfo{pages}{4527} (\bibinfo{year}{1987}).

\bibitem[{\citenamefont{Evans et~al.}(1993)\citenamefont{Evans, Henderson,
  Hoyle, Parry, and Sabeur}}]{Evans1993}
\bibinfo{author}{\bibfnamefont{R.}~\bibnamefont{Evans}},
  \bibinfo{author}{\bibfnamefont{J.~R.} \bibnamefont{Henderson}},
  \bibinfo{author}{\bibfnamefont{D.~C.} \bibnamefont{Hoyle}},
  \bibinfo{author}{\bibfnamefont{A.~O.} \bibnamefont{Parry}}, \bibnamefont{and}
  \bibinfo{author}{\bibfnamefont{Z.~A.} \bibnamefont{Sabeur}},
  \bibinfo{journal}{Mol. Phys.} \textbf{\bibinfo{volume}{80}},
  \bibinfo{pages}{755} (\bibinfo{year}{1993}).

\bibitem[{\citenamefont{Evans and Parry}(1989)}]{Evans1989}
\bibinfo{author}{\bibfnamefont{R.}~\bibnamefont{Evans}} \bibnamefont{and}
  \bibinfo{author}{\bibfnamefont{A.~O.} \bibnamefont{Parry}},
  \bibinfo{journal}{J. Phys.: Condens. Matter} \textbf{\bibinfo{volume}{1}},
  \bibinfo{pages}{7207} (\bibinfo{year}{1989}).

\bibitem[{\citenamefont{Evans et~al.}(1992)\citenamefont{Evans, Hoyle, and
  Parry}}]{Evans1992}
\bibinfo{author}{\bibfnamefont{R.}~\bibnamefont{Evans}},
  \bibinfo{author}{\bibfnamefont{D.~C.} \bibnamefont{Hoyle}}, \bibnamefont{and}
  \bibinfo{author}{\bibfnamefont{A.~O.} \bibnamefont{Parry}},
  \bibinfo{journal}{Phys. Rev. A} \textbf{\bibinfo{volume}{45}},
  \bibinfo{pages}{3823} (\bibinfo{year}{1992}).

\bibitem[{\citenamefont{Fern{\'a}ndez et~al.}(2013)\citenamefont{Fern{\'a}ndez,
  Chac{\'o}n, Tarazona, Parry, and Rasc{\'o}n}}]{Fernandez2013}
\bibinfo{author}{\bibfnamefont{E.~M.} \bibnamefont{Fern{\'a}ndez}},
  \bibinfo{author}{\bibfnamefont{E.}~\bibnamefont{Chac{\'o}n}},
  \bibinfo{author}{\bibfnamefont{P.}~\bibnamefont{Tarazona}},
  \bibinfo{author}{\bibfnamefont{A.~O.} \bibnamefont{Parry}}, \bibnamefont{and}
  \bibinfo{author}{\bibfnamefont{C.}~\bibnamefont{Rasc{\'o}n}},
  \bibinfo{journal}{Phys. Rev. Lett.} \textbf{\bibinfo{volume}{111}},
  \bibinfo{pages}{096104} (\bibinfo{year}{2013}).

\bibitem[{\citenamefont{Evans}(1992)}]{EvansDFT1992}
\bibinfo{author}{\bibfnamefont{R.}~\bibnamefont{Evans}}, in
  \emph{\bibinfo{booktitle}{{ Fundamentals of Inhomogeneous Fluids}}}, edited
  by \bibinfo{editor}{\bibfnamefont{D.}~\bibnamefont{Henderson}}
  (\bibinfo{publisher}{Marcel {D}ekker, {I}nc.}, \bibinfo{year}{1992}),
  chap.~\bibinfo{chapter}{3}.

\bibitem[{\citenamefont{Das and Binder}(2011)}]{Das2011}
\bibinfo{author}{\bibfnamefont{S.~K.} \bibnamefont{Das}} \bibnamefont{and}
  \bibinfo{author}{\bibfnamefont{K.}~\bibnamefont{Binder}},
  \bibinfo{journal}{Phys. Rev. Lett.} \textbf{\bibinfo{volume}{107}},
  \bibinfo{pages}{235702} (\bibinfo{year}{2011}).

\bibitem[{\citenamefont{Fisher and Jin}(1991)}]{Fisher1991}
\bibinfo{author}{\bibfnamefont{M.~E.} \bibnamefont{Fisher}} \bibnamefont{and}
  \bibinfo{author}{\bibfnamefont{A.~J.} \bibnamefont{Jin}},
  \bibinfo{journal}{Phys. Rev. B} \textbf{\bibinfo{volume}{44}},
  \bibinfo{pages}{1430} (\bibinfo{year}{1991}).

\bibitem[{\citenamefont{Lu et~al.}(1985)\citenamefont{Lu, Evans, and {Telo da
  Gama}}}]{Lu1985}
\bibinfo{author}{\bibfnamefont{B.~Q.} \bibnamefont{Lu}},
  \bibinfo{author}{\bibfnamefont{R.}~\bibnamefont{Evans}}, \bibnamefont{and}
  \bibinfo{author}{\bibfnamefont{M.~M.} \bibnamefont{{Telo da Gama}}},
  \bibinfo{journal}{Mol. Phys.} \textbf{\bibinfo{volume}{55}},
  \bibinfo{pages}{1319} (\bibinfo{year}{1985}).

\bibitem[{\citenamefont{Evans}(1981)}]{Evans1981}
\bibinfo{author}{\bibfnamefont{R.}~\bibnamefont{Evans}}, \bibinfo{journal}{Mol.
  Phys.} \textbf{\bibinfo{volume}{42}}, \bibinfo{pages}{1169}
  (\bibinfo{year}{1981}).

\bibitem[{\citenamefont{Fisk and Widom}(1969)}]{Fisk1969}
\bibinfo{author}{\bibfnamefont{S.}~\bibnamefont{Fisk}} \bibnamefont{and}
  \bibinfo{author}{\bibfnamefont{B.}~\bibnamefont{Widom}}, \bibinfo{journal}{J.
  Chem. Phys.} \textbf{\bibinfo{volume}{50}}, \bibinfo{pages}{3219}
  (\bibinfo{year}{1969}).

\bibitem[{\citenamefont{Sullivan}(1979)}]{Sullivan1979}
\bibinfo{author}{\bibfnamefont{D.~E.} \bibnamefont{Sullivan}},
  \bibinfo{journal}{Phys. Rev. B} \textbf{\bibinfo{volume}{20}},
  \bibinfo{pages}{3991} (\bibinfo{year}{1979}).

\bibitem[{\citenamefont{Sullivan}(1981)}]{Sullivan1981}
\bibinfo{author}{\bibfnamefont{D.~E.} \bibnamefont{Sullivan}},
  \bibinfo{journal}{J. Chem. Phys.} \textbf{\bibinfo{volume}{74}},
  \bibinfo{pages}{2604} (\bibinfo{year}{1981}).

\bibitem[{\citenamefont{Tarazona and Evans}(1983)}]{Tarazona1983}
\bibinfo{author}{\bibfnamefont{P.}~\bibnamefont{Tarazona}} \bibnamefont{and}
  \bibinfo{author}{\bibfnamefont{R.}~\bibnamefont{Evans}},
  \bibinfo{journal}{Mol. Phys.} \textbf{\bibinfo{volume}{48}},
  \bibinfo{pages}{799} (\bibinfo{year}{1983}).

\bibitem[{\citenamefont{Parry and Evans}(1988)}]{Parry1988}
\bibinfo{author}{\bibfnamefont{A.~O.} \bibnamefont{Parry}} \bibnamefont{and}
  \bibinfo{author}{\bibfnamefont{R.}~\bibnamefont{Evans}},
  \bibinfo{journal}{Mol. Phys.} \textbf{\bibinfo{volume}{65}},
  \bibinfo{pages}{455} (\bibinfo{year}{1988}).

\bibitem[{\citenamefont{H{\"o}fling}()}]{Hpc}
\bibinfo{author}{\bibfnamefont{F.}~\bibnamefont{H{\"o}fling}},
  \bibinfo{note}{{P}rivate {C}ommunication}.

\bibitem[{\citenamefont{Ohta and Kawasaki}(1977)}]{Ohta1977}
\bibinfo{author}{\bibfnamefont{T.}~\bibnamefont{Ohta}} \bibnamefont{and}
  \bibinfo{author}{\bibfnamefont{K.}~\bibnamefont{Kawasaki}},
  \bibinfo{journal}{Prog. Theor. Phys.} \textbf{\bibinfo{volume}{58}},
  \bibinfo{pages}{467} (\bibinfo{year}{1977}).

\end{thebibliography}

\end{document}